\documentclass[a4paper,11pt]{article}

\usepackage{jcappub} 

\usepackage{siunitx}
\usepackage{graphicx}
\usepackage{hyperref}
\usepackage{caption}
\usepackage{subcaption}
\usepackage{amsmath,amssymb,wasysym}
\usepackage{booktabs}
\usepackage{amsfonts}
\usepackage{physics}
\usepackage[nameinlink,noabbrev]{cleveref}

\DeclareSIUnit\permille{\text{\textperthousand}}

\title{Sensitivity of a closed dielectric haloscope to axion dark matter}

\collaborationImg{\includegraphics{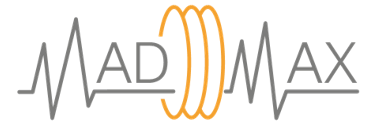}}

\author[a]{A.~Ivanov}
\author[b,1]{D.~Leppla-Weber\note{Corresponding author}}
\author[c]{B.~Ary dos Santos Garcia}
\author[c]{D.~Bergermann}
\author[a]{H.~Byun}
\author[a]{A.~Caldwell}
\author[d]{V.~Dabhi}
\author[d]{C.~Diaconu}
\author[a]{J.~Diehl}
\author[a]{G.~Dvali}
\author[a]{B.~Döbrich}
\author[b]{J.~Egge}
\author[e]{E.~Garutti}
\author[f]{S.~Heyminck}
\author[g]{T.~Houdy}
\author[d]{F.~Hubaut}
\author[h]{J.~Jochum}
\author[a]{A.~Kazemipour}
\author[g]{Y.~Kermaidic}
\author[i]{S.~Knirck}
\author[f]{M.~Kramer}
\author[a]{D.~Kreikemeyer-Lorenzo}
\author[e]{C.~Krieger}
\author[a]{C.~Lee}
\author[a]{X.~Li}
\author[b]{A.~Lindner}
\author[a]{B.~Majorovits}
\author[a]{J.~Maldonado}
\author[b]{A.~Martini}
\author[g]{A.~Miyazaki}
\author[c]{E.~Öz}
\author[d]{P.~Pralavorio}
\author[a]{G.~Raffelt}
\author[j]{J.~Redondo}
\author[b]{A.~Ringwald}
\author[b]{J.~Schaffran}
\author[c]{A.~Schmidt}
\author[e]{L.~Stankewitz}
\author[a]{F.~Steffen}
\author[h]{C.~Strandhagen}
\author[h]{I.~Usherov}
\author[c]{H.~Wang}
\author[f]{G.~Wieching}

\affiliation[a]{Max-Planck-Institut f{\"ur} Physik, Garching, Germany}
\affiliation[b]{Deutsches Elektronen-Synchrotron DESY, Germany}
\affiliation[c]{III. Physikalisches Institut A,  RWTH Aachen University, Aachen, Germany}
\affiliation[d]{Aix Marseille Univ, CNRS/IN2P3, CPPM, Marseille, France}
\affiliation[e]{Universit\"at Hamburg, Hamburg, Germany}
\affiliation[f]{Max-Planck-Institut f{\"ur} Radioastronomie, Bonn, Germany}
\affiliation[g]{Université Paris-Saclay, CNRS/IN2P3, IJCLab, Orsay, France}
\affiliation[h]{Physikalisches Institut, Eberhard Karls Universit{\"a}t T{\"u}bingen, T{\"u}bingen, Germany}
\affiliation[i]{Fermi National Accelerator Laboratory, Batavia, IL, USA}
\affiliation[j]{Universidad Zaragoza, Zaragoza, Spain}

\abstract{%
We present a method to determine the sensitivity of a closed dielectric haloscope to axion dark matter.
Dielectric haloscopes aim to probe the theoretically well-motivated axion mass range of $\sim\SI{26}{\micro\eV}$ to $\sim\SI{500}{\micro\eV}$ by utilizing a stack of dielectric disks and a mirror to enhance the axion-photon conversion within an external magnetic field.
Their conversion volume is nearly axion-mass independent, thereby favoring large-scale designs to increase sensitivity.
The large volume causes simulations to be computationally expensive and time-consuming. 
This paper presents a simple model that can be used to determine the sensitivity of the experiment with minimal computational resources.
The model is able to describe the electromagnetic response of a closed dielectric haloscope, accounting for realistic geometric imperfections, as
well as the noise introduced by the receiver system.
It is applied to data taken with a MAgnetized Disk and Mirror Axion Experiment (MADMAX) prototype within the 1.6 T Morpurgo magnet at CERN.
This work underpins the first axion dark matter search using a dielectric haloscope and provides the foundation for future dark matter searches with MADMAX.
}

\begin{document}

\maketitle
\flushbottom

\section{Introduction}

In the search for lightweight dark matter (DM), cavity haloscope experiments~\cite{Sikivie:1983ip} have been an essential tool for many years now.
They operate under the assumption that the DM from the galactic halo converts to photons.
Haloscope experiments were originally motivated by the axion, predicted by a solution to the strong CP problem~\cite{PhysRevLett.38.1440,PhysRevD.16.1791,PhysRevLett.40.223}.
Nevertheless, they are also sensitive to other non-relativistic, wavelike DM candidates, such as the dark photon (DP).

By including axions in the Standard Model, Maxwell's equations receive an additional current like source term $\mathbf{J}_a = g_{a\gamma}\dot a \mathbf{B}_e$ composed of the axion-photon coupling $g_{a\gamma}$, the time derivative of the axion field $\dot a$ and the external magnetic field $\mathbf{B}_e$.
Similarly, the DP results in an effective current $\mathbf{J}_\mathrm{dp} = \chi\dot{\mathbf{E}}_\chi$ with the kinetic mixing angle $\chi$ and the time derivative of the DP electric field $\dot{\mathbf{E}}_\chi$.
The momentum of non-relativistic DM particles can be neglected, leading to a direct relationship between the particle mass $m_\mathrm{dm}$ and the resulting photon frequency $\nu \simeq m_\mathrm{dm}c^2/h$.
The measurable signal power of a haloscope $P_\mathrm{sig}$ is given by the overlap between the effective DM current $\mathbf{J}_\mathrm{dm} = \mathbf{J}_\text{a/dp}$ and the electric field $\mathbf{E}$ coupled to the detector: $P_\mathrm{sig} \propto \int dV \mathbf{E} \cdot \mathbf{J}_\mathrm{dm}$~\cite{Egge:2022gfp}.
Due to the macroscopic de Broglie wavelength of the considered DM particles $\lambda_\text{db}$ of $\mathcal{O}(\SI{10}{\m})$, $\mathbf{J}_\text{dm}$ is approximately uniform over the extent of the experiment.
Typically, resonant radio frequency (RF) cavities operated at the fundamental transverse magnetic mode are used to maximize $\mathbf{E}$ and therefore $P_\text{sig}$.
Because cavity size at the fundamental mode decreases with frequency, higher DM masses lead to smaller volumes and lower signal power, limiting sensitivity.
Different approaches exist to overcome this limitation~\cite{alpha,horns_2013}, one being the dielectric haloscope~\cite{Caldwell:2016dcw}.

\begin{figure}
    \centering
    \includegraphics[width=0.80\textwidth]{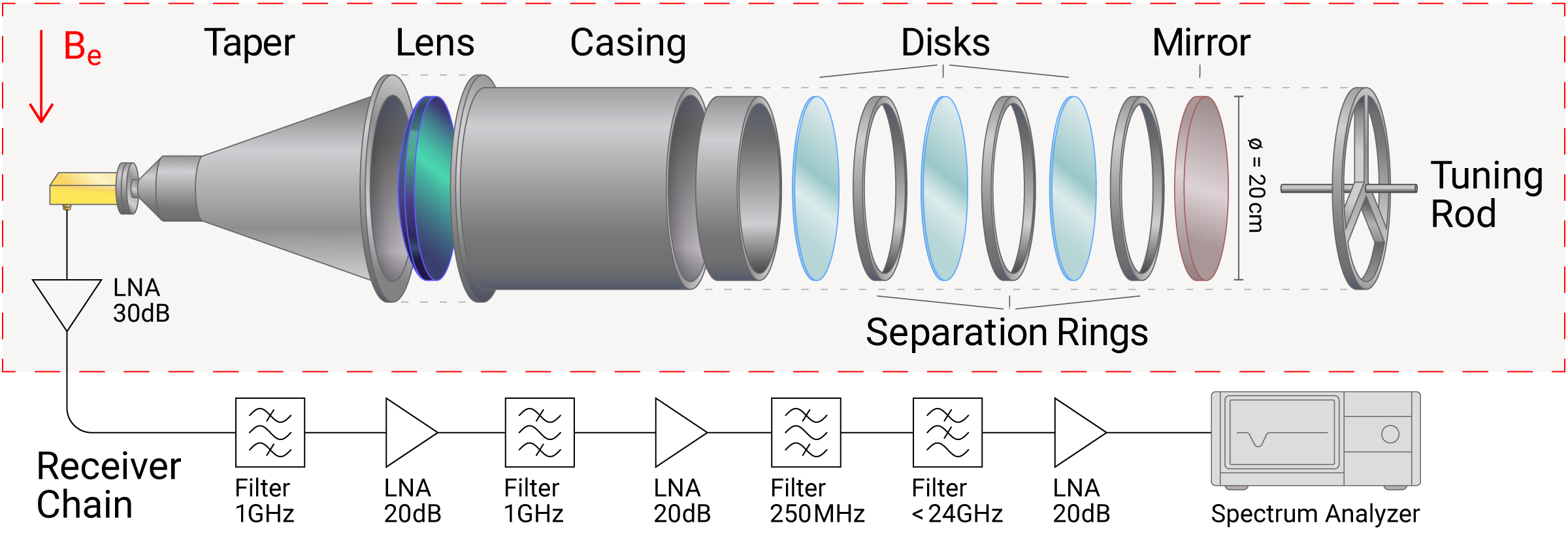}
    \caption{Exploded schematic view of the closed dielectric booster CB200 and the receiver chain considered in this paper. In addition to the disks and mirror, placed in an aluminum casing, it includes a taper and a dielectric lens designed to maximize coupling of the axion signal power to the low-noise receiver chain.
    The tuning rod allows to apply pressure to the mirror, therefore slightly offsetting its position.
    The shaded region indicates the components exposed to the magnetic field $\mathbf{B}_e$. The setup is identical to the latest axion search performed at CERN's Morpurgo magnet and the figure taken from its publication~\cite{ary_dos_santos_garcia_first_2025}.}
    \label{fig:MADMAX}
\end{figure}

The MAgnetized Disk and Mirror Axion eXperiment
(MADMAX)~\cite{Jaeckel:2013eha,Caldwell:2016dcw, MADMAX:2019pub} is such a dielectric haloscope designed to access the mass range $\SI{40}{\micro\eV} \lesssim m_\text{dm} \lesssim \SI{400}{\micro\eV}$, corresponding to a photon wavelength of $\SIrange{3}{30}{\mm}$.
Based on the magnetized mirror axion DM search idea~\cite{horns_2013}, it uses a
\textit{booster} composed of a set of parallel dielectric disks and a metallic mirror, as shown in \cref{fig:MADMAX}, to enhance a potential axion DM signal.
Depending on the specific booster boundaries, the detector couples either to the fundamental transverse-electric or fundamental transverse-electromagnetic mode, polarized parallel to the disks and mirror in direction of the magnetic field.
This in principle allows for an arbitrary transverse scaling of disk and mirror diameter to increase the conversion volume.
In practice it is however limited by mechanical constraints and machining tolerances.
The placement of multiple dielectric disks allows to shape the distribution of the electric field $\mathbf{E}$ in the longitudinal direction, further increasing $\int dV \mathbf{E}\cdot\mathbf{J}_\text{dm}$ and therefore $P_\text{sig}$.
This effectively decouples the conversion volume from the photon wavelength.
The resulting signal power $P_\text{sig}$ can be equivalently described by coherent photon emissions at the interfaces between air and disk regions, allowing for constructive interference and resonances to increase the experiments sensitivity~\cite{Millar_2017}.
In relation to the power emitted by a magnetized mirror, $P_0$, $P_\text{sig}$ is enhanced by the frequency dependent boost factor $\beta^2 := P_\text{sig}/P_0$.
By adjusting the disk and mirror positions its center frequency and width can be tuned.

The sensitivity of a dielectric haloscope to axions is expressed as the axion-photon coupling $g_{a\gamma}$ needed to produce a given signal-to-noise ratio (SNR) within the integration time $\Delta t$ for the local axion DM density~$\rho_a$~\cite{Millar_2017}:
\begin{equation}
    \label{eq:sensitivity}
    \begin{split}
        |g_{a\gamma}| = &\;\SI{3.5e-11}{\giga\eV^{-1}} \sqrt{\frac{2\times 10^3}{{\beta^2}}} \sqrt{\frac{T_\mathrm{sys}}{\SI{300}{K}}} \\
        & \times\left( \frac{\SI{0.1}{m}}{r}\right) \left( \frac{\SI{1}{T}}{B_e} \right) \left( \frac{\SI{2.2}{days}}{\Delta t}\right)^{1/4} \!\!\sqrt{\frac{\mathrm{SNR}}{5}} \\
        & \times\left( \frac{m_a}{\SI{80}{\micro\eV}}\right)^{5/4}
        \sqrt{\frac{\SI{0.3}{\giga\eV/cm^3}}{\rho_a}}\;.
    \end{split}
\end{equation}
Most of the quantities governing $|g_{a\gamma}|$ are either explicitly set by the experimentalist or directly measurable.
This includes the target axion mass $m_a$, the radius of the disks $r$ and the external magnetic field $\mathbf{B_e} = \hat{y}B_e$, aligned parallel to the disks and mirrors.
After performing a standard power calibration~\cite{pozar2011microwave},  described in more detail in \cref{sec:power-cal}, power spectra taken with the receiver system can be converted to the system temperature $T_\text{sys}$.
The only remaining parameter is therefore the boost factor $\beta^2$.
Its determination requires a more complex procedure since a direct calibration would require to replicate the DM current density $\mathbf{J}_\text{dm}$ exactly, which is experimentally challenging.
Furthermore, a full-scale simulation of the entire setup is computationally demanding, due to its large domain size  of $>\mathcal{O}(10^3\lambda^3)$ relative to the photon wavelength $\lambda$.
This stands in contrast to typical cavity domain sizes of $\mathcal{O}(10\lambda^3)$.

The MADMAX Collaboration has recently published first lower limits on the coupling of axion DM to photons $g_{a\gamma}$ and the kinetic mixing angle $\chi$ utilizing two different types of dielectric haloscopes that allow for different approaches to determine $\beta^2$.
An \textit{open} booster with three \(\diameter\SI{300}{\mm}\) free-standing sapphire disks was used to perform the first MADMAX DP DM search and improved existing limits by nearly three orders of magnitude~\cite{MADMAX:2024jnp}.
In this case, the procedure to determine $\beta^2$ was based on a direct measurement of the antenna excited electric field $\mathbf{E}$ within the booster, from which $\beta^2$ can be calculated~\cite{Egge:2023cos}.
The measurement of the electric field requires the insertion of a perturbing object into the booster structure and is therefore only applicable to a setup with free standing disks with no surrounding boundary, to which we refer to as an \textit{open} booster.
Determining $\beta^2$ solely by electric field measurements is very time consuming, which can be improved by relying in parts on a model or simulation.

In a more recent work, a \textit{closed} booster consisting of three $\diameter \SI{200}{mm}$ sapphire disks encased in an aluminum cylinder, referred to as CB200 and depicted in \cref{fig:MADMAX}, marked the first axion DM search~\cite{ary_dos_santos_garcia_first_2025} with a dielectric haloscope.
CB200 uses a dielectric lens and taper to couple the receiver system to the fundamental transverse-electric mode of the booster.
The procedure developed for this search only requires access to the port of the taper, relying on a complete model of the booster and connected receiver system, which is the topic of this article.
The fixed boundary conditions provided by the aluminum cylinder make it a \textit{closed} booster and ease modeling.
As opposed to the open booster, only a finite set of modes propagates within such a system.
The search consisted of five periods of data taking, referred to as physics-runs, performed using two different booster configurations in which separation rings of different widths were used.
Between physics-runs of the same configuration, the frequency of maximal sensitivity was adjusted by $\mathcal{O}(\SI{10}{MHz})$ using the tuning rod, depicted in \cref{fig:MADMAX}, to press against the mirror.

This paper presents the analysis that was performed to obtain the already published results~\cite{ary_dos_santos_garcia_first_2025} in full detail.
In \cref{sec:method}, a booster and receiver chain model is described and exemplary fitted to the data of the first physics run.
With the resulting parameters, the boost factor of the used setup can be calculated.
The model is then thoroughly validated against full-wave finite-element-method (FEM) simulations in \cref{sec:model-verification} and applied to the physics-runs in \cref{sec:data-taking}, focusing on the resulting uncertainty and time stability of the system.
In \cref{sec:conclusion} we provide our conclusions on the performed study.

\section{Booster and receiver noise model}
\label{sec:method}

The frequency dependent boost factor $\beta^2$ of a booster is defined by the configuration of the dielectric disks, depending on their positions and permittivities.
It is sensitive to $\mathcal{O}(\si{\micro\m})$ deviations of not just the disk and mirror positions, but also additional imperfections such as tiny gaps or misalignments.
Full-wave simulations for which all of those parameters need to be implemented are very time consuming.
Apart from the general computational effort of simulating such large-scale systems, it would also require fine-tuning the parameters describing the imperfections and geometric details to reproduce the exact behavior of the system, multiplying the computational power required for an exact description.
Instead, an effective model is used to reproduce the reflectivity of the booster $\Gamma(\nu) = \frac{\phi_\mathrm{out}}{\phi_\mathrm{in}}$ , determined by externally exciting the system by a known field amplitude $\phi_\mathrm{in}$ and measuring the reflected field amplitude $\phi_\mathrm{out}$ using a Vector Network Analyzer (VNA).
The model has to be simple enough to computationally allow for fitting of its multiple parameters.
In \cref{sec:booster} such a model based on network theory~\cite{pozar2011microwave} is presented.
By fitting its parameters it is able to reproduce the measured reflectivity of the booster $\Gamma(\nu)$ and can then be used to calculate the corresponding $\beta^2(\nu)$.
The accuracy of this calculation is discussed in \cref{sec:model-verification}.

When taking data for the DM search, the booster is connected to the receiver system via a $\SI{50}{\ohm}$ coaxial connection.
Because the receiver system is not perfectly matched to that connection, it reflects a small fraction of the potential signal, altering the boost factor.
With the receiver connected, it is only possible to measure system temperature spectra, consisting of the noise emitted by the receiver and booster.
A noise model reproducing such a spectrum is therefore necessary to quantify the effect of a mismatched receiver system on the boost factor.
The noise model of the receiver is described in \cref{sec:rec-noise}, requiring additional system temperature measurements of known standards to determine its parameters.
It is then combined with the booster model in \cref{sec:noise-full} to reproduce the measured system temperature of the booster by fitting the only free parameter left: the electrical length that connects the receiver to the booster.

In the following, the full procedure to determine the sensitivity of a given CB200 configuration is demonstrated for the first physics-run published in~\cite{ary_dos_santos_garcia_first_2025}, referred to as run 1.1.
It consists of determining:
\begin{enumerate}
    \item booster model parameters by fitting to a reflectivity measurement of the booster,
    \item receiver noise model parameters by fitting to system temperature measurements of known standards,
    \item the electrical distance connecting both models by fitting to a system temperature measurement of the booster connected to the receiver system.
\end{enumerate}

\subsection{Booster model}
\label{sec:booster}

We first focus on modeling the booster’s reflectivity to extract the parameters necessary to calculate $\beta^2$ from a reflectivity measurement.
Due to its conducting cylindrical boundaries, wave propagation within the closed booster is well described by the modes of a cylindrical waveguide.
In the presence of a homogeneous magnetic field, the axion-induced effective current $\mathbf{J}_a$ excites mainly the fundamental transverse electric (TE$_{11}$) mode, which propagates along the booster axis, orthogonal to the mirror surface.
We call the resonance of this mode the \textit{booster mode}.
Such guided wave propagation is modeled using a well-known transmission line (TL) approach \cite{pozar2011microwave}, implemented in Advanced Design System (ADS)~\cite{ads}.
It is equivalent to using a transfer matrix based method as described in~\cite{Millar_2017}, but enables easy integration of the booster into the noise model, also implemented in ADS and described in \cref{sec:rec-noise} and \cref{sec:noise-full}.

The model does not consider higher-order mode contributions to the boost factor.
It is therefore a fundamental requirement to identify a frequency range in which only the booster mode is present.
This requirement is studied in \cref{sec:higher-order-modes} and \cref{sec:beadpull}, using simulations and complementary measurements.
 
\begin{figure}
    \centering
    \includegraphics[width=1\textwidth]{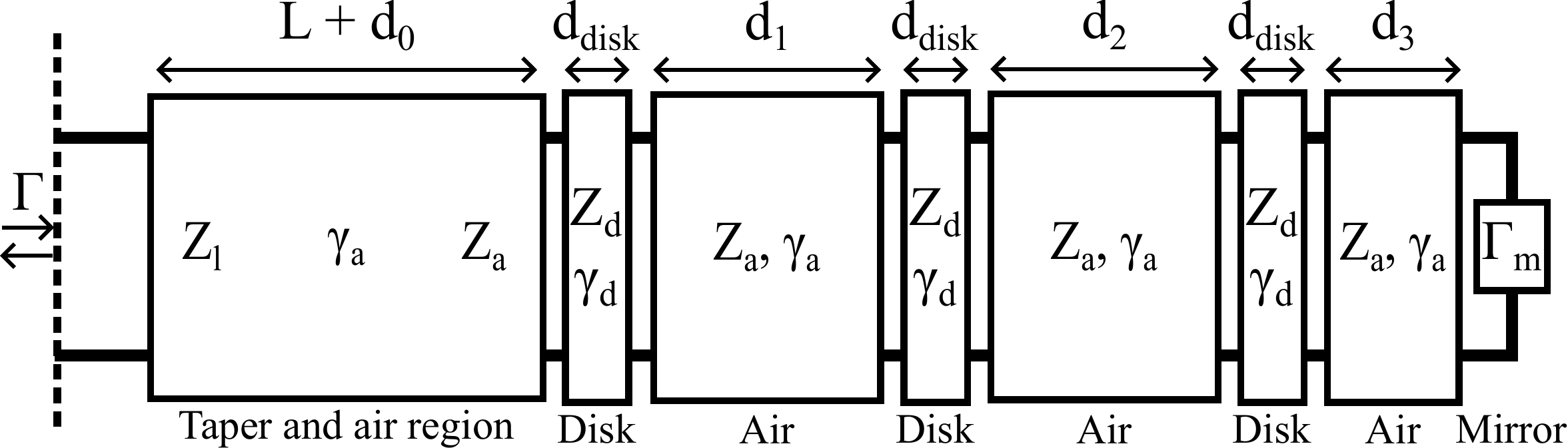} 
    \caption{Transmission line configuration used to model the booster. It consists of four air and three disk regions of respective lenghts $d_i$ and $d_\text{disk}$, defined by their corresponding impedances $Z_{a}$ and $Z_{d}$ as well as propagation constants $\gamma_a$ and $\gamma_d$.
    The leftmost air region also includes the length of the taper $L$.
    It transitions from $Z_a$ to the coaxial line impedance $Z_l = \SI{50}{\ohm}$.
    The mirror is modeled by its reflectivity $\Gamma_m$ given by its finite conductivity.
    The reflectivity measurement is modeled at the reference plane indicated by the dashed line, which corresponds to the connection of the VNA to the taper, the booster input. }
    \label{fig:TL}
\end{figure}

\Cref{fig:TL} shows the booster model, composed of TL segments for the disk (d) and air (a) regions, characterized by their impedances $Z_{d}$ and $Z_{a}$, respectively, their propagation constants $\gamma_d$ and $\gamma_a$ and their individual lengths $d_\text{disk}$ and $d_{i}$ ($i=0,1,2,3$).
The TL segments directly correspond to the booster parts as shown in \cref{fig:MADMAX}.
While the disks are all of $\SI{1}{\mm}$ thickness, the separation rings are chosen individually for the different configurations.
The electric field of the resonating booster mode is mostly concentrated between the mirror and the closest disk, such that it is most sensitive to the distance between the two, $d_3$.
The leftmost region includes the measured length of the taper $L=\SI{166 \pm 1}{\mm}$ and the width of the leftmost separation ring $d_0 = \SI{37.40 \pm 0.01}{\mm}$ for configuration 1 and $d_0 = \SI{21.14 \pm 0.01}{\mm}$ for configuration 2.
The region's length does not affect the boost factor and is therefore kept fixed for each configuration.
The taper is assumed to provide a perfect coupling between its coaxial port and the booster mode, implemented by an ideal taper that transitions from the coaxial line impedance $Z_l = \SI{50}{\ohm}$ to the air impedance $Z_a$ without introducing reflections.
This assumption is verified in \cref{sec:higher-order-modes}.

The equations in this paragraph are based on the description of a circular waveguide in~\cite{pozar2011microwave}, unless stated otherwise.
The impedance as a function of the relative permittivity $\epsilon$ in the case of the circular waveguide is given by:
\begin{equation}
	Z(\epsilon) = Z_0 \sqrt{1/\epsilon} \frac{k(\epsilon)}{\gamma(\epsilon)},
\end{equation}
with the free space impedance $Z_0 \simeq \SI{377}{\ohm}$, dispersion relation $k(\epsilon) = \omega \sqrt{\epsilon}/c$, angular frequency $\omega = 2\pi\nu$ and speed of light in vacuum $c$.
The disk and air impedances  $Z_d = Z(\epsilon_d)$ and $Z_a=Z(\epsilon_a)$ differ between air and disk regions only due to their different relative permittivities: $\epsilon_d = 9.36$ for sapphire and $\epsilon_a = 1$ for air.
The change in phase per unit length along the booster axis is defined by the propagation constant of the TE$_{11}$ mode:
\begin{equation}
	\gamma(\epsilon) = \sqrt{\frac{\omega^2\epsilon}{c^2} -
		\frac{{p'_{11}}^2}{r^2}}
\end{equation}
with waveguide radius $r$,  and the first root of the derivative of the Bessel function of order one $p'_{11} = 1.841$. 

Loss due to the finite conductivity $\sigma_\text{Al} = \SI{3.77e7}{\siemens\per\m}$ for the aluminum booster walls is described by the attenuation per unit distance:
\begin{equation}
	\alpha_c = \frac{1}{rkZ\gamma}\sqrt{\frac{\omega\mu_0}{2\sigma_\text{Al}}}\left(\frac{{p'_{11}}^2}{r^2} + \frac{k^2}{{p'_{11}}^2 -
		1}\right).
\end{equation}
Additional dielectric loss in the disk segments is accounted for by:
\begin{equation}
 \alpha_d =  \frac{\tan \delta}{2} \, \frac{k(\epsilon_d)^2}{\gamma(\epsilon_d)},
\end{equation}
where a uniformly filled lossy dielectric with loss tangent $\tan\delta$ is considered.
Dielectric loss within air regions is neglected.
The non-perfect reflection $\Gamma_m$ at the aluminum mirror due to its finite conductivity is given by $\Gamma_m = 1 - 2\sqrt{2 \omega \epsilon_0 / \sigma_\mathrm{Al}}$~\cite{hummel_electronic_2011}.
By connecting transmission line segments as shown in \cref{fig:TL}, with propagation constants $\gamma$ and attenuation $\alpha_c + \alpha_d$ for disks and $\alpha_c$ for air within ADS, an ADS S-parameter simulation is used to calculate the resulting sum of reflections at the interfaces and therefore the reflectivity of the booster $\Gamma(\nu) = S_{11}$.

\begin{figure}
    \centering
    \includegraphics[width=0.8\linewidth]{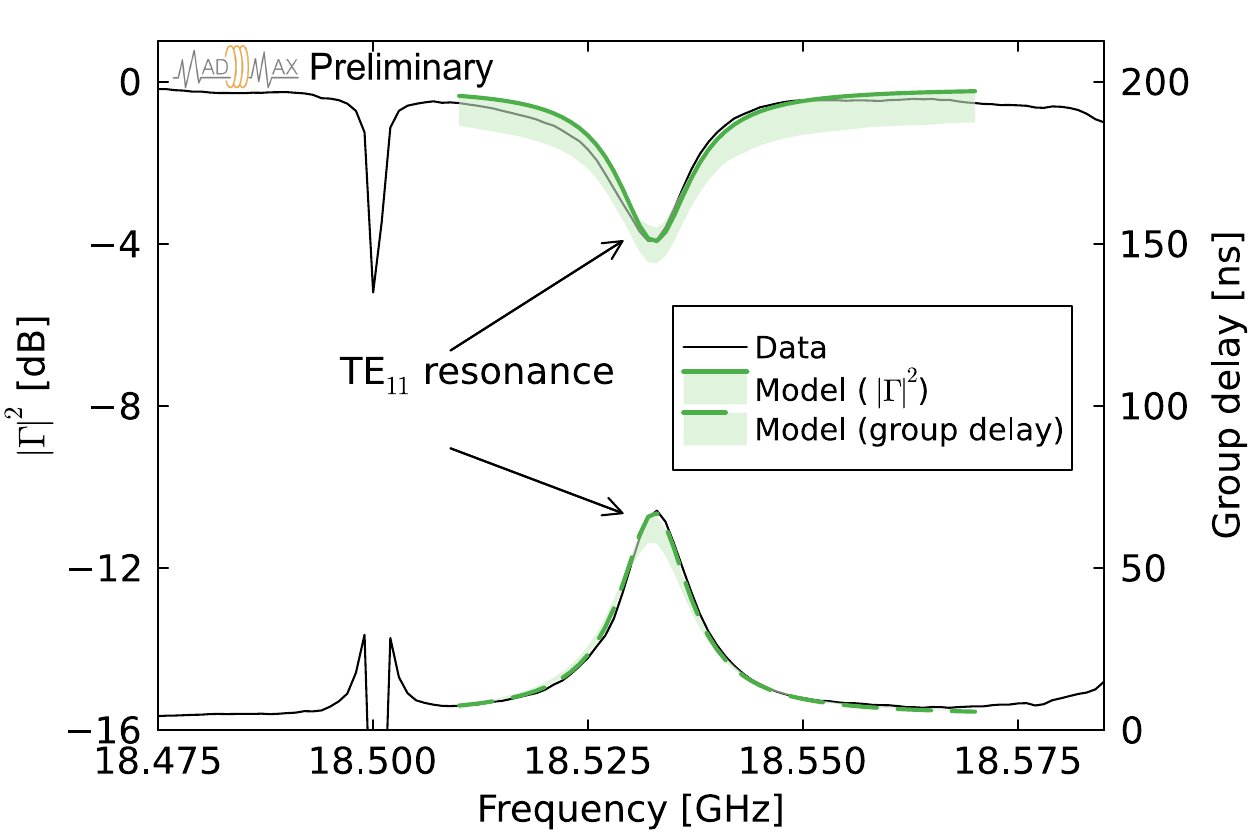}
    \caption{Measured (black) and modeled (green) reflectivity of CB200, run 1.1 (see \cref{tab:booster_model}), after parameter fit. Top curves: reflected power $|\Gamma|^2$. Bottom curves: group delay $\tau_\text{gd}$. The model curve shows the best fit and the band the $1\sigma$ fit uncertainty within the frequency range of the booster mode.
    The TE$_{11}$ resonance, that is the booster mode, is visible as a dip in reflected power and a peak in group delay.
    Fit parameters are listed in \cref{tab:booster_model} and fixed parameters in \cref{tab:fixed-params} in the \cref{sec:params} .}
    \label{fig:reflfit-example}
\end{figure}

By fitting the disk spacings $d_i$, as well as disk permittivity $\epsilon_d$, thickness $d_\text{disk}$ and loss $\tan\delta$ of the described model, the booster mode visible in the measured reflectivity of the CB200 booster is reproduced within uncertainties, as can be seen in \cref{fig:reflfit-example}.
The model and measurement agree for both the reflected power $|\Gamma|^2$ (top curves) and group delay $\tau_\text{gd} = -\frac{\partial\arg{\Gamma}}{\partial\omega}$ (bottom curves).
Since the refractive index of the disks $n = \sqrt{\epsilon_d}$ and the disk thickness $d_\text{disk}$ are almost fully correlated, the product $nd_\text{disk}$, referred to as disk phase thickness, is fitted instead of each separately.
As initial parameters for the fit, the physical dimensions of the disks and separation rings are used as well as the expected permittivity $\epsilon_d = 9.36$ and dielectric loss $\tan\delta = 10^{-5}$ of sapphire.
The fitted parameters, together with their uncertainties, will be discussed in \cref{sec:data-taking}.
It also contains \cref{tab:booster_model}, which includes the parameters used in the fit exemplary presented here for run 1.1.
All fixed parameters are summarized in \cref{tab:fixed-params} in the \cref{sec:params}.

To calculate the frequency dependent boost factor, the axion-induced excitation is modeled by including voltage sources of power $P = P_0 (1 - 1 /
\epsilon_d)$ with opposite phase to each side of every disk and one of $P = P_0$ to the
mirror, with $P_0$ being the power emitted by a magnetized mirror~\cite{horns_2013}.  The signal power $P_\text{sig}$ is then given by an ADS power probe at the input of the booster, resulting in $\beta_\text{1D}^2 = P_\text{sig}/P_0$.
This quantity does not yet take into account the transverse field shape within the booster.
While the effective current $\mathbf{J}_a$ due to the axion within the magnetic field is approximately constant over the whole volume, the tangential part of the electric field of the booster mode $\mathbf{E}_\mathrm{TE_{11}}$ vanishes at the metallic boundaries.

The coupling is therefore reduced by the transverse geometric overlap of the two, resulting in
$\beta^2 = |\eta_A|^2 \beta_\text{1D}^2$ with
\begin{align}
	\label{eq:overlap}
	|\eta_A|^2 &= \frac{\left|\int_A dA\ \mathbf{E}_\mathrm{TE_{11}}\cdot\mathbf{J}_a\right|^2}
	{\int_A dA\ |\mathbf{E}_\mathrm{TE_{11}}|^2 \int_A dA\ |\mathbf{J}_a|^2}\bigg\rvert_{\mathbf{J}_a=\text{const}} \\
	&= \frac{\left|\int_A dA\ \mathbf{E}_\mathrm{TE_{11}}\cdot
		\hat y\right|^2}{A\int_A dA\ |\mathbf{E}_\mathrm{TE_{11}}|^2} = 0.84.
\end{align}
This quantity is conceptually similar to the form factor of a cavity, with the difference that the form factor
describes the overlap of the resonating mode with the axion current over the volume of the
cavity while $|\eta_A|^2$ only considers the transverse overlap, since the longitudinal behavior is
already described by the booster model.

\subsection{Receiver noise model}
\label{sec:rec-noise}

To take into account the effects of the receiver system on the boost factor, we need to model the system temperature of the combined system of booster and receiver.
An ADS noise simulation is used for this purpose.
In our case, the dominant noise sources are the first-stage low noise amplifier (LNA) and the booster itself.
While ADS automatically simulates noise of passive devices such as the booster according to their signal attenuation, active devices such as the LNA require the addition of specific noise sources.

\begin{figure}
    \centering
    \includegraphics[width=0.8\linewidth]{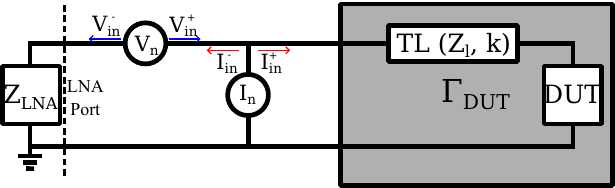}
    \caption{Schematic of the circuit used to model LNA noise. It consists of a transmission line of impedance $Z_l = \SI{50}{\ohm}$ terminated by the LNA impedance $Z_\text{LNA}$ on one side and a device-under-test (DUT) such as the booster on the other side. The LNA noise is introduced by a voltage and current noise source connected in parallel. The dashed line represents the LNA input port.}
    \label{fig:lna-model}
\end{figure}

We model the amplifier noise with a correlated current and voltage noise
source, $I_n$ and $V_n$, connected in parallel~\cite{linearnoise}, as depicted in~\cref{fig:lna-model}.
The circuit is terminated by the first stage LNA, implemented as a one-port device with the impedance set to the frequency-dependent impedance of the input port of the LNA $Z_\text{LNA}$, measured by a VNA.
The device-under-test (DUT) can be described either by its measured or modeled reflectivity $\Gamma_\text{DUT}$.
The system temperature as measured with the receiver system is then modeled using the simulated noise power delivered to the one-port device, representing the LNA.
\begin{figure}
    \centering
    \includegraphics[width=0.8\linewidth]{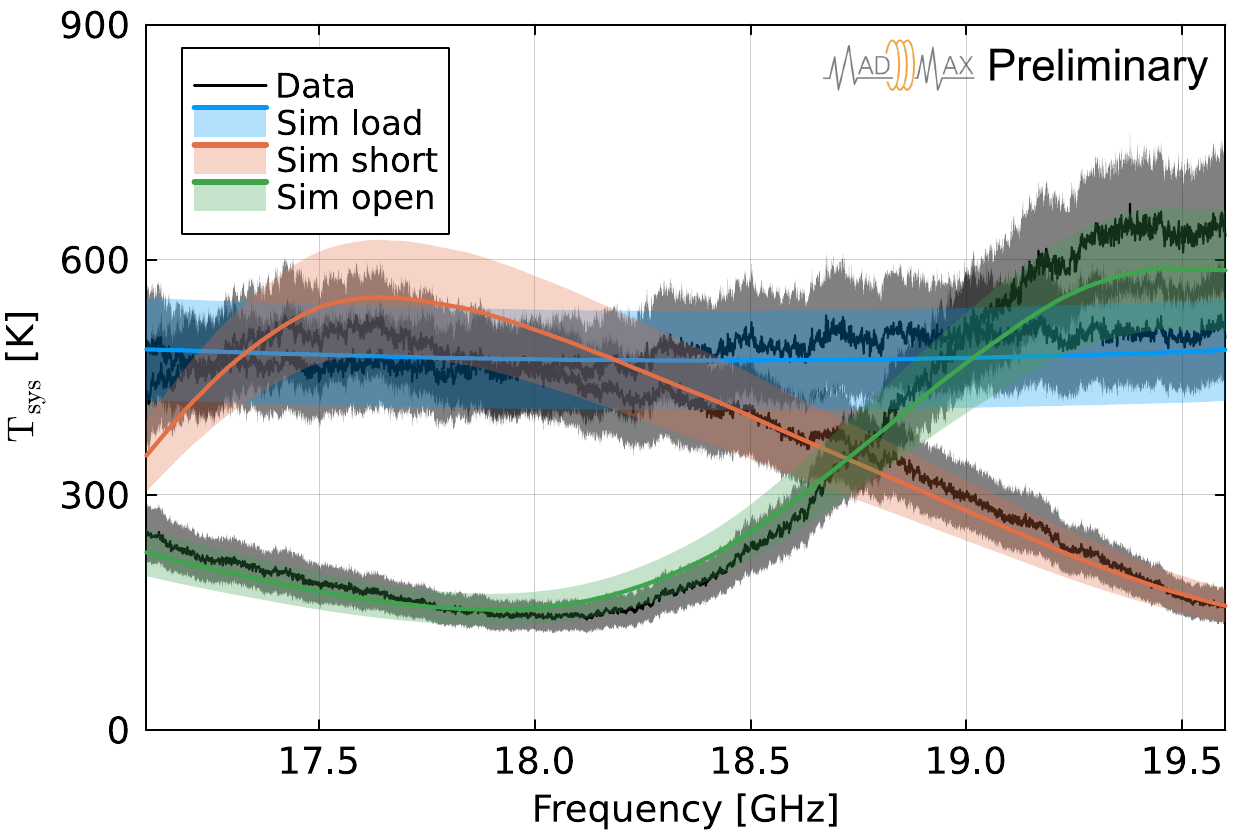}
    \caption{Measured and simulated noise spectra $T_\text{sys}$ of the amplifier connected to open, short and matched load standards.
    The bands around the curves show their associated uncertainty, stemming from the power calibration of the data.
    The load (blue) is close to constant over the frequency range, while the open (green) and short (orange) measurements exhibit an oscillation of opposite phase. The simulation matches the measurements within uncertainties.}
    \label{fig:lna-standards}
\end{figure}
Due to the potential correlation of $V_n$ and $I_n$, this results in four parameters for the receiver noise model: voltage noise amplitude $V_n$, current noise amplitude $I_n$, correlation magnitude $|c|$ and correlation angle $\phi_c$.
They are extracted from fits to system temperature spectra, that are measured with the LNA connected to a short, open and matched load standard as the DUT.
$\Gamma_\text{DUT}$ is given by their idealized reflectivities $\Gamma_\text{short} = -1$, $\Gamma_\text{open} = 1$ and $\Gamma_\text{load} = 0$.
For the open and short standards, the distance between LNA and standard is added as an additional fit parameter.
The parameters are fitted to all three spectra simultaneously, with the result shown in \cref{fig:lna-standards} and the fitted parameters listed in \cref{tab:noise-model} in the \cref{sec:params}.

The noise model reproduces the measurements within their uncertainties, stemming from the receiver power calibration~(\cref{sec:power-cal}).
It matches nicely the slightly skewed shape of the oscillation present in open and short measurements.
The maximum of the spectrum of the short standard, at around \SI{17.5}{\GHz}, is slightly overestimated and the maximum of the spectrum of the open standard, at around $\SI{19.4}{\GHz}$, is slightly underestimated.
This hints at an increase in noise emitted by the LNA with frequency, but is well within the uncertainties that will be discussed in \cref{sec:data-taking}.

\subsection{Combined noise model}
\label{sec:noise-full}

\begin{figure}
    \centering
    \includegraphics[width=\linewidth]{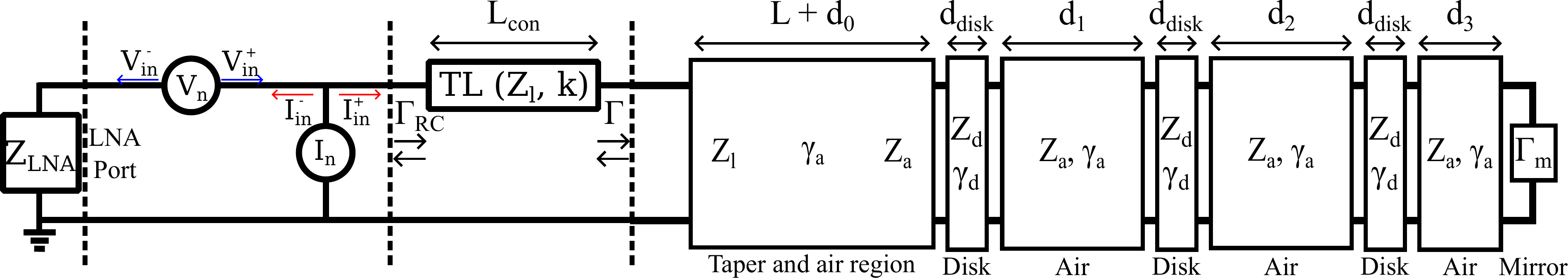}
    \caption{Combined noise model consisting of the receiver noise model and booster model connected by a transmission line of length $L_\text{con}$. The dashed lines represent the reference planes from which the booster and receiver chain reflectivity measurements, $\Gamma$ and $\Gamma_\text{RC}$ respectively, are taken with the VNA and the LNA input port.}
    \label{fig:combined-model}
\end{figure}

To combine the LNA and booster model, they are connected by a transmission line of length $L_\text{con}$ within ADS.
This results in the combined model as shown in \cref{fig:combined-model}, where the device under test from \cref{fig:lna-model} is replaced by the booster model described in \cref{sec:booster}.
Due to the impedance mismatch of the LNA to the line, the boost factor as simulated by ADS differs from the simulated one without the LNA.
\begin{figure}
    \centering
    \includegraphics[width=0.8\linewidth]{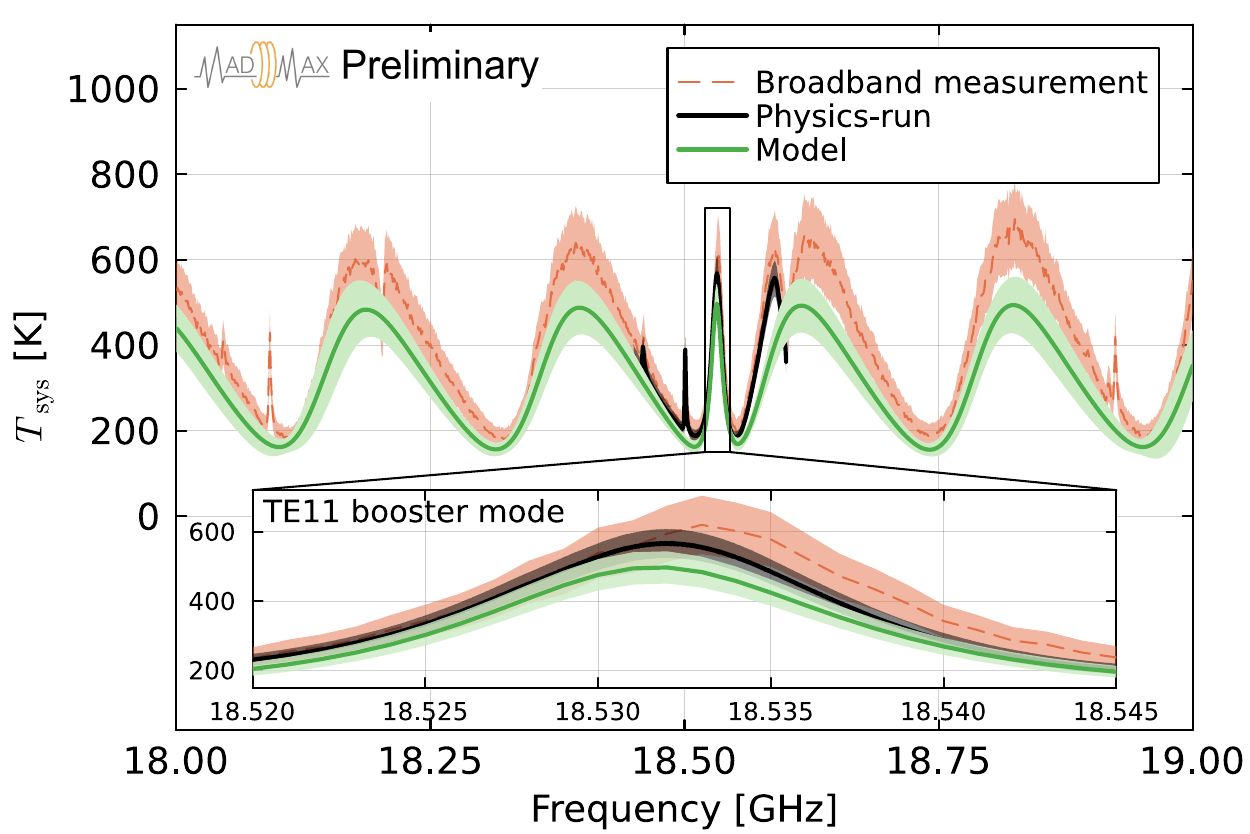}
    \caption{Measured and modeled power spectra of CB200, resonant at $\sim\SI{18.53}{\GHz}$ for run~1.1 (see \cref{tab:booster_model}). The model (green solid) reproduces the positions of minima and maxima of the broadband oscillation present in the calibration measurement (orange dashed) as well as the booster mode frequency of the physics-run (black solid).
    The amplitude is consistently slightly underestimated.
    The inset displays a zoom in on the frequency of the booster mode.}
    \label{fig:noisefit-example}
\end{figure}
To quantify the change, the only additionally required parameter is the length $L_\text{con}$ of the transmission line connecting LNA and booster.

\Cref{fig:noisefit-example} shows a broadband measurement (\SI{1}{\GHz}) of the system temperature of CB200 (orange dashed), the first narrowband spectrum taken during run 1.1 (black) as well as the reproduction by the noise model (green solid).

The broadband measurement features a very prominent oscillation, which is the result of a standing wave between the LNA and the booster.
The distance between maxima and minima in frequency space directly depends on the remaining unknown parameter $L_\text{con}$.
By fitting $L_\text{con}$, the positions of minima and maxima over the full \SI{1}{\GHz} range of the measurement are consistently reproduced.
Having a broadband measurement increases the precision on $L_\text{con}$, since multiple minima and maxima are visible.
This configuration of the receiver system is dedicated to determining the boost factor before the physics-run.

The physics-run spectrum was taken in a narrowband configuration of the receiver system optimized for the axion search.
It does not include multiple minima and maxima of the broadband oscillation, but instead allows to spot any changes in booster mode frequency between the time of the broadband measurement and the start of the physics-run.
They are visible as a shift in frequency of the feature corresponding to the booster mode, magnified in the inset.
In the shown case, the booster mode frequency slightly changed, necessitating an adjustment of the model.
As will be further elaborated on in \cref{sec:bf-det}, the change is caused by an initially unstable mirror position.
It is therefore accommodated for by slightly varying the distance between mirror and closest disk $d_3$ of the model.
In this specific case, the required change was only $\sim\SI{1}{\micro\m}$, corresponding to a $\sim\SI{1}{\MHz}$ shift in frequency.
Uncertainties will be discussed in \cref{sec:data-taking}.

The model does not include higher order modes and therefore does not reproduce additional peaks and dips present in the spectra outside of the booster mode frequency.
Furthermore, the amplitudes between broadband measurement and physics-run can change slightly because of the reconfiguration of the receiver system as well as environmental factors such as the temperature.
This also affects the amplitude obtained from the model, since it is defined from the previously performed fits to the standards.
An overall deviation in amplitude, however, does not affect the boost factor calculation at all, since the parameter extracted from this measurement, $L_\text{con}$, only changes the accurately reproduced positions of minima and maxima.

\section{Model verification}
\label{sec:model-verification}
While the presented booster model is able to reproduce the performed reflectivity measurements within uncertainties, it remains to be shown, that the resulting boost factor calculation holds up to scrutiny.
We therefore present multiple full-wave FEM simulations to verify that the booster model accurately describes the axion induced excitation and is able to accommodate for 3D effects such as tilts and disk deformation using effective parameters.
Those FEM simulations do not include the receiver part of the setup, since it is fully single-mode and therefore completely described by the presented transmission line based receiver model.
Previous studies have already been performed on the effect of geometrical inaccuracies of disks on the boost factor of an open booster setup with $\geq\ 20$ disks \cite{MADMAX:2021lxf}, as opposed to the three disk closed booster considered in this work.

In \cref{sec:comsol-ideal} and \cref{sec:3d-effects} the booster model is compared to those simulations, determining its limits and accuracy.
In \cref{sec:higher-order-modes} and \cref{sec:beadpull} it is then described how the booster mode is identified and isolated from unwanted modes, which is an inherent requirement for the validity of the booster model.

\subsection{Ideal case}
\label{sec:comsol-ideal}
\begin{figure}
    \centering
    \begin{subfigure}[t]{\textwidth}
        \centering
    \includegraphics[width=0.8\textwidth]{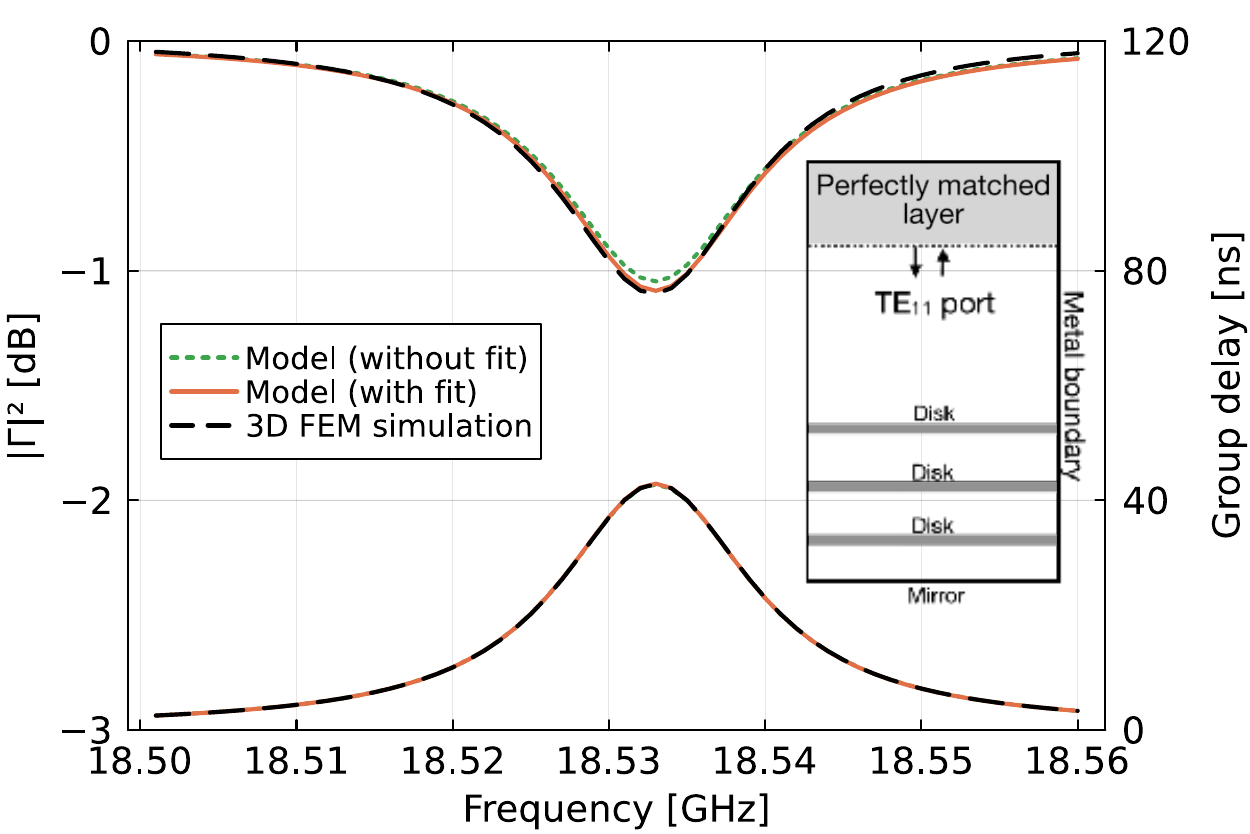}
        \caption{Comparison of reflectivities with schematic of the simulation setup in the inset.}
        \label{fig:cmp-ideal-a}
    \end{subfigure}\\
    \vspace*{0.5em}
    \begin{subfigure}[t]{\textwidth}
        \centering
        \hspace*{-1.3cm}\includegraphics[width=0.74\textwidth]{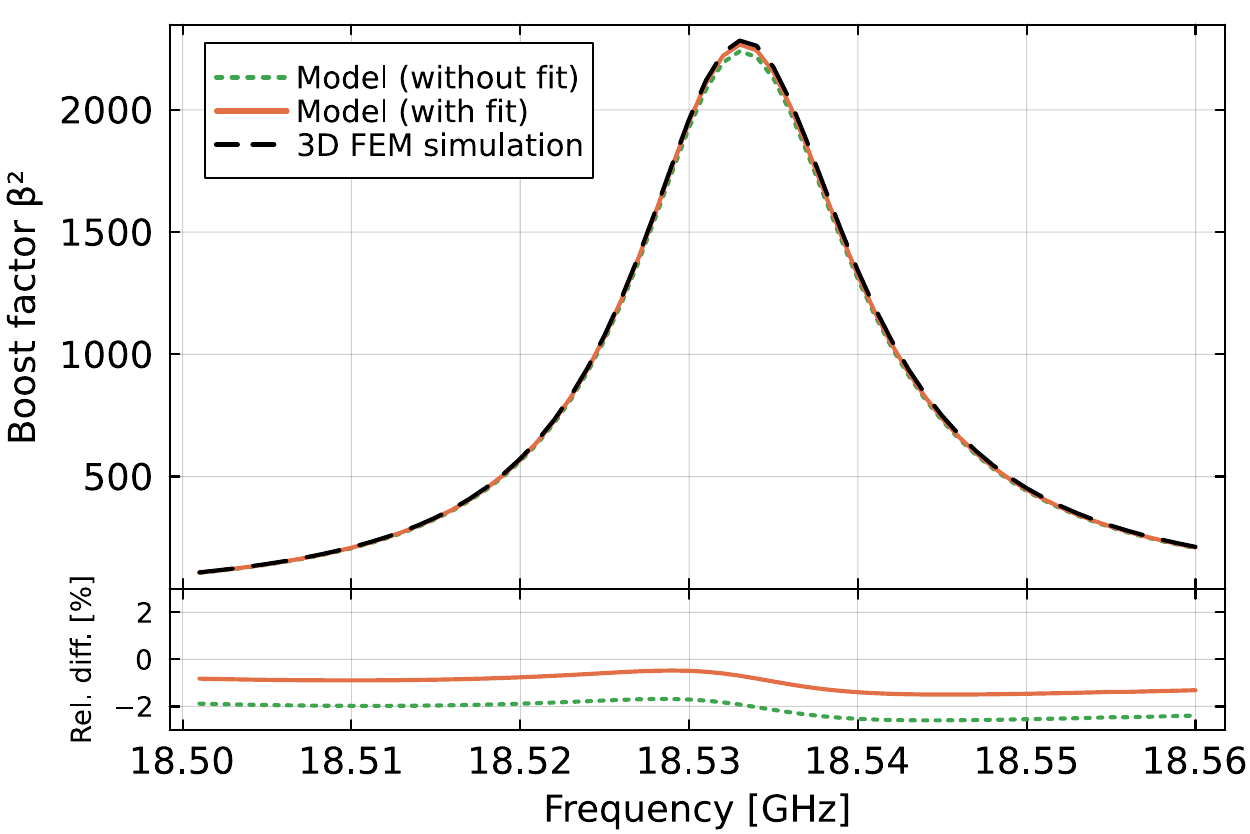}
        \caption{Comparison of boost factors and their relative difference.}
        \label{fig:cmp-ideal-b}
    \end{subfigure}
    \caption{Comparison of the reflectivity and boost factor obtained from the booster model and the 3D FEM simulation of the setup sketched in the inset of (a) (rotated by $90^\circ$ for graphical reasons).
    The spectra of reflected power $|\Gamma|^2$ and group delay $\tau_\text{gd}$ are shown in (a), whereas the boost factor spectra $\beta^2$ together with their relative difference are shown in (b).
    The results obtained from the booster model without fitting its parameters (green dotted) and with fitting its parameters (orange solid) match the results obtained with a 3D FEM simulation (black dashed) of an ideal booster comparable to CB200 during run~1.1 within $\SI{3}{\percent}$.
    }
    \label{fig:cmp-ideal}
\end{figure}

We first consider an ideal booster with perfectly planar disks and a mirror, free of tilts or misalignment, surrounded by a highly conducting boundary and equipped with an ideal, non-reflecting taper supporting only the intended TE$_{11}$ mode.
As sketched in the inset of \cref{fig:cmp-ideal}, such a taper is simulated using a perfectly matched layer behind the port.
This idealized case is implemented in a full-wave benchmark FEM simulation performed using COMSOL Multiphysics\textsuperscript{\textregistered}~\cite{comsol}.
A realistic taper, exciting higher-order modes (HOMs), will be discussed in \cref{sec:higher-order-modes}.
Due to computational constraints, the radius of the cylinder is set to $\SI{5}{\cm}$ instead of $\SI{10}{\cm}$. Since an ideal booster does not excite HOMs, the results are independent of booster radius.
The axion-induced excitation is modeled with a background current given by $\mathbf{J}_a$, aligned to the polarization of the TE$_{11}$ mode.
The simulated boost factor $\beta^2$ is calculated at the TE$_{11}$ port from the power coupled to the TE$_{11}$ mode.
External excitation is implemented with a standard TE$_{11}$ port, from which the simulated reflectivity $\Gamma(\nu)$ is obtained.

In \cref{fig:cmp-ideal} we show the 3D FEM simulation result for an ideal booster with three sapphire disks, that is resonant at $\sim\SI{18.55}{\GHz}$, similar to the booster in run 1.1 as considered in \cref{sec:method}.
The comparison to the simulated reflectivity is performed with and without parameter fitting (using exactly the simulated material and geometric parameters).

Already without a fit, the group delay obtained with the booster model is indistinguishable from the one obtained with the 3D FEM simulation.
Only the reflected power deviates slightly, which is likely caused by a different treatment of conductive losses at the boundaries and the mirror between the booster model and the COMSOL 3D FEM simulation.
In terms of boost factor $\beta^2$, an excellent agreement is seen between the model and 3D FEM simulation with $<\SI{3}{\percent}$ difference throughout the frequency spectrum without fitting the parameters and $<\SI{2}{\percent}$ difference after the fit.
These results are consistent across different configurations and demonstrate the capability of the booster model to accurately calculate the boost factor of an ideal booster and already show how the model can make up for slight deviations by adjusting its effective parameters.

\subsection{Treatment of 3D effects}
\label{sec:3d-effects}
The booster model used to describe the electromagnetic response of the booster is inherently one-dimensional.
Realistic inhomogeneities of material properties, such as the dielectric constant, conductivity or thickness, break the azimuthal symmetry of the geometry and redistribute the field.
This can lead to a change in boost factor due to:
(i) deformation of the intended booster mode and coupling to HOMs;
(ii) shift of frequency of the booster mode, also resulting in changes of the reflectivity; and
(iii) localized field deviations that alter the coupling to the input antenna.
Such three-dimensional effects are accounted for in the booster model through effective parameters.
This section demonstrates that the booster model is capable to describe the small discrepancies typically observed between idealized and measured booster responses by allowing the model parameters to deviate from their nominal physical values.

Mirror tilt is the dominant 3D effect on the booster response as the booster is set up for a resonance between mirror and first disk.
In practice, the disk spacings are well controlled by precision-machined metal rings, while the mirror position is deliberately left adjustable to enable fine-tuning of the frequency of maximal sensitivity.
This design choice provides a useful degree of tunability, but also modifies the booster response.
Accordingly the influence of the mirror boundary deserves closer examination.

\begin{figure}[htbp]
    \centering
    \begin{subfigure}[t]{\textwidth}
        \centering
        \includegraphics[width=0.8\textwidth]{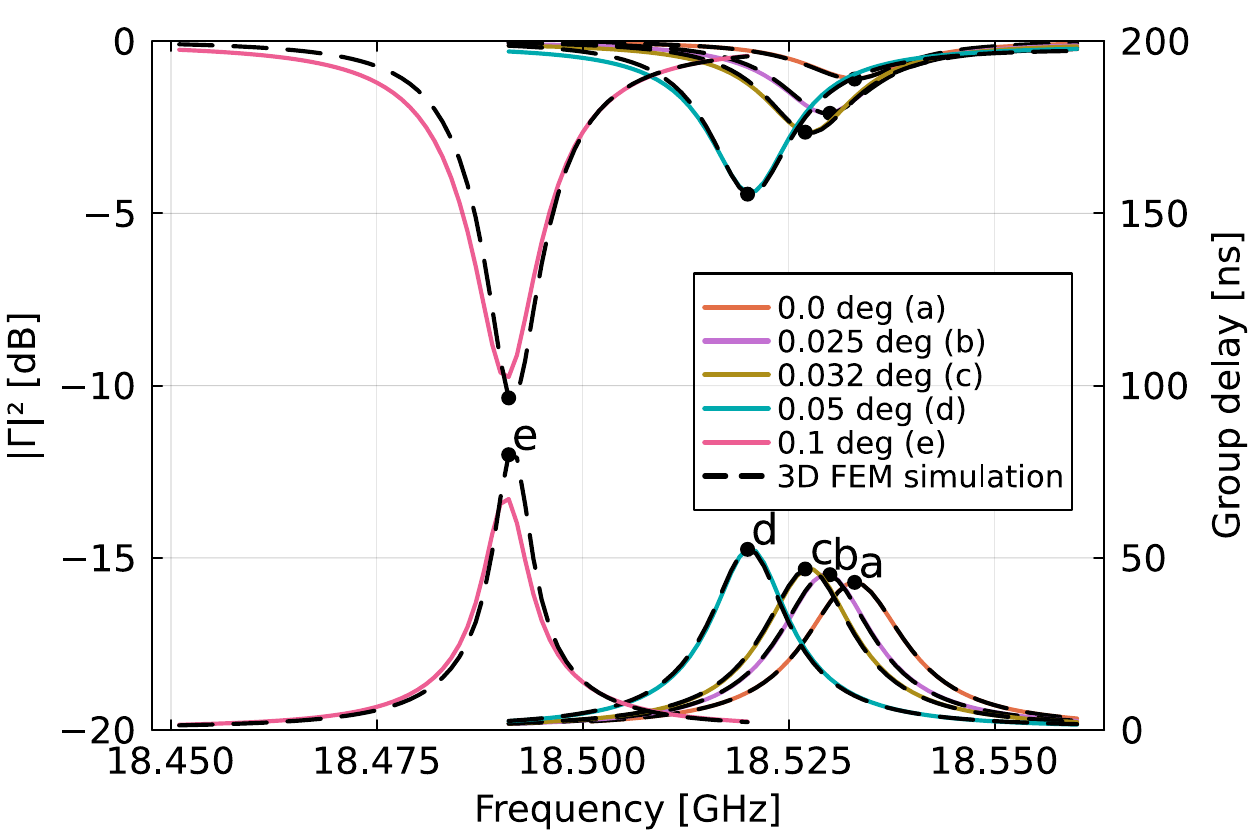}
        \caption{Comparison of reflectivity from 3D FEM simulation and booster model.}
        \label{fig:S11_vs_boost_tilt-a}
    \end{subfigure}\\
    \vspace*{0.5em}
    \begin{subfigure}[t]{\textwidth}
        \centering\hspace*{-1.3cm}\includegraphics[width=0.74\textwidth]{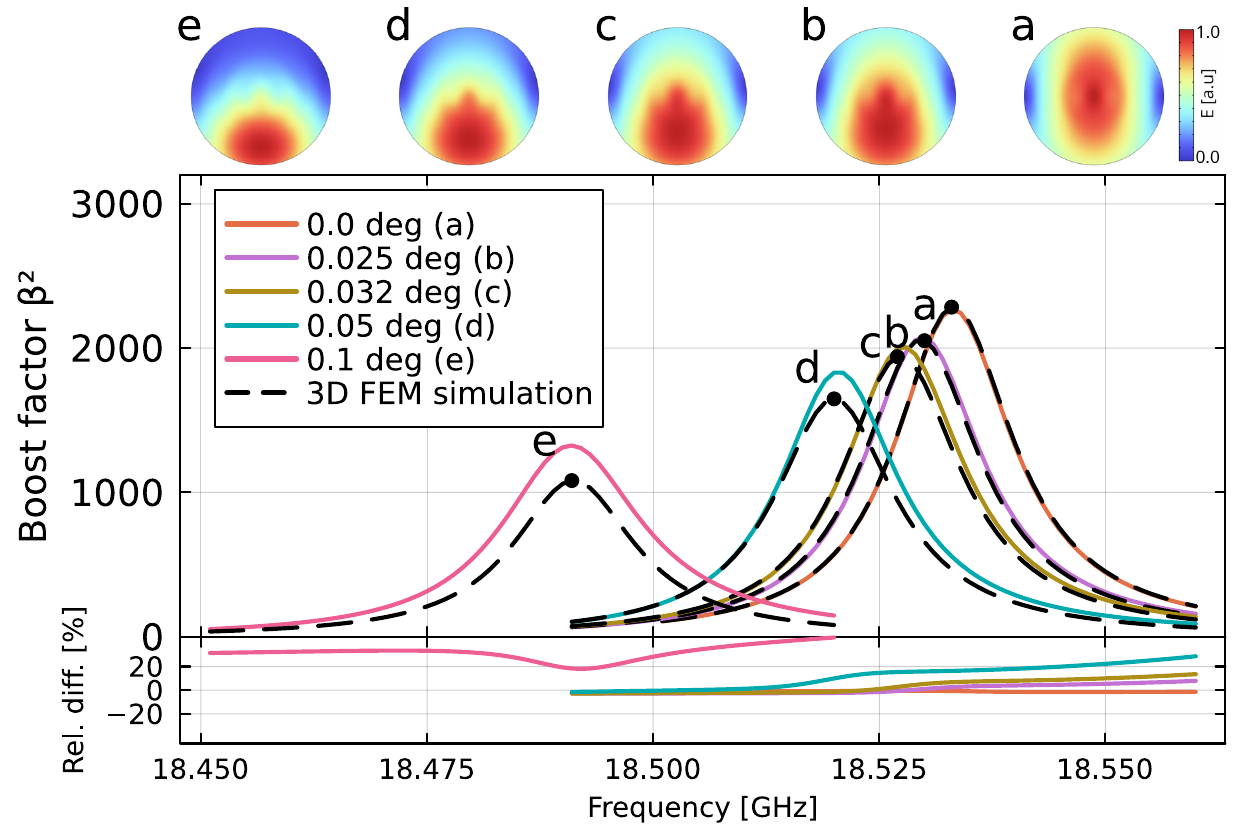}
        \caption{Comparison of boost factor from 3D FEM simulation and booster model.}
    \label{fig:S11_vs_boost_tilt-b}
    \end{subfigure}
    \caption{Simulated (black dashed) and modeled (different colors solid) booster response under mirror tilt around the $x$-axis.
    The effective model parameters are fit to reproduce the reflectivities obtained from a 3D FEM simulation including mirror tilts from \SIrange{0}{0.1}{\deg}.
    The spectra of reflected power $|\Gamma|^2$ and group delay $\tau_\text{gd}$ are shown in (a), whereas the boost factor spectra $\beta^2$ together with their relative difference are shown in (b).
    At the top of (b), the simulated transverse electric field distribution at the TE$_{11}$ port is shown for the frequencies of maximum boost factor.
    }
    \label{fig:S11_vs_boost_tilt}
\end{figure}

To realistically account for the effect of the mirror tilt, we modify the 3D FEM simulation described in \cref{sec:comsol-ideal} to include a tilted mirror, defined by a rotation around a given axis.
The $x$- and $y$-axis define the plane of the untilted mirror, with $\mathbf{J}_a$ polarized along the $y$-axis and the origin at the mirror center.
The $z$-axis is the normal of that plane, pointing towards the taper.

A tilt of the mirror perturbs the field, such that the power in the booster mode reduces due to mixing with unwanted HOMs.
The boost factor evaluation only considers the power of the booster mode, excluding power contribution from HOMs.
This provides a conservative estimate of the power reaching the detector, since in practice the taper is able to convert some of the HOM power back into the desired mode.
The reduced booster radius of \SI{5}{\cm} remains very large compared to the wavelength of $\sim\SI{1.6}{\cm}$ and therefore does not significantly affect this analysis: a radius of \SI{5}{\cm} supports roughly 100 propagating modes at \SI{18.5}{\GHz}, allowing for the field perturbations expected from the small deviations from an ideal booster that we simulate.

\Cref{fig:S11_vs_boost_tilt} shows the simulated booster response (black dashed) for a mirror tilt between \SI{0}{\deg} (case a) and \SI{0.1}{\deg} (case e) around the $x$-axis.

As expected, increasing tilt produces an observable change in the booster mode, which is likewise reflected in the boost factor behavior.
The results indicate:
(i) the booster mode shifts to lower frequency, as seen from the displacement of the corresponding feature in the reflectivity and the shift of the boost factor line;
(ii) the dip in reflected power broadens and deepens progressively as the tilt increases;
(iii) the boost factor broadens and decreases correspondingly,
(iv) the axion-induced field distribution is distorted and departs from the ideal TE$_{11}$ shape (case a).

The final step is to fit the booster model to the tilt simulation and derive the boost factor.
The procedure described in \cref{sec:booster} is applied to reflectivity spectra from the 3D FEM simulation, that is disk positions $d_i$, disk phase thickness $nd_\text{disk}$ and dielectric loss $\tan\delta$ of the booster model are fitted to match the simulated reflectivities.
The boost factor that is computed from the fitted booster model is then compared to the simulation.

The results from the fitted model are overlayed with the simulated response in \cref{fig:S11_vs_boost_tilt} (different colors, solid). 
The booster model reproduces the simulated reflectivities remarkably well in all cases except for the largest tilt of \SI{0.1}{\deg}.
The modeled boost factor closely follows the frequency behavior of the simulation.
Only the amplitude is increasingly overestimated for higher tilts.
For a \SI{0.05}{\deg} tilt, which is compatible with the worst case of the runs later analyzed, the model still agrees with the 3D FEM simulation within 10 \% at the boost factor maximum.
At \SI{0.1}{\deg}, the boost factor maximum is overestimated by $\sim\SI{20}{\percent}$.
This behavior mainly arises from the degraded coupling to the axion current $\mathbf{J}_a$ not captured fully by the booster model: while the booster model uses its effective dielectric loss parameter to accurately reproduce the coupling visible in the reflected power $|\Gamma|^2$ as an increasingly deep dip, it underestimates the effect of the distorted electric field.
This distortion, and hence decrease in overlap, limits the power that couples to the receiver and therefore reduces $\beta^2$.
This is a purely 3D effect beyond the scope of the booster model and demonstrates its limitations at large tilts.
It was further checked that additionally rotating the tilted mirror around the $z$-axis

yields consistent results with the presented tilt orientation showing the highest deviation from the simulated boost factor.

\begin{figure}
    \centering
    \begin{subfigure}[t]{\textwidth}
        \centering
        \includegraphics[width=0.8\textwidth]{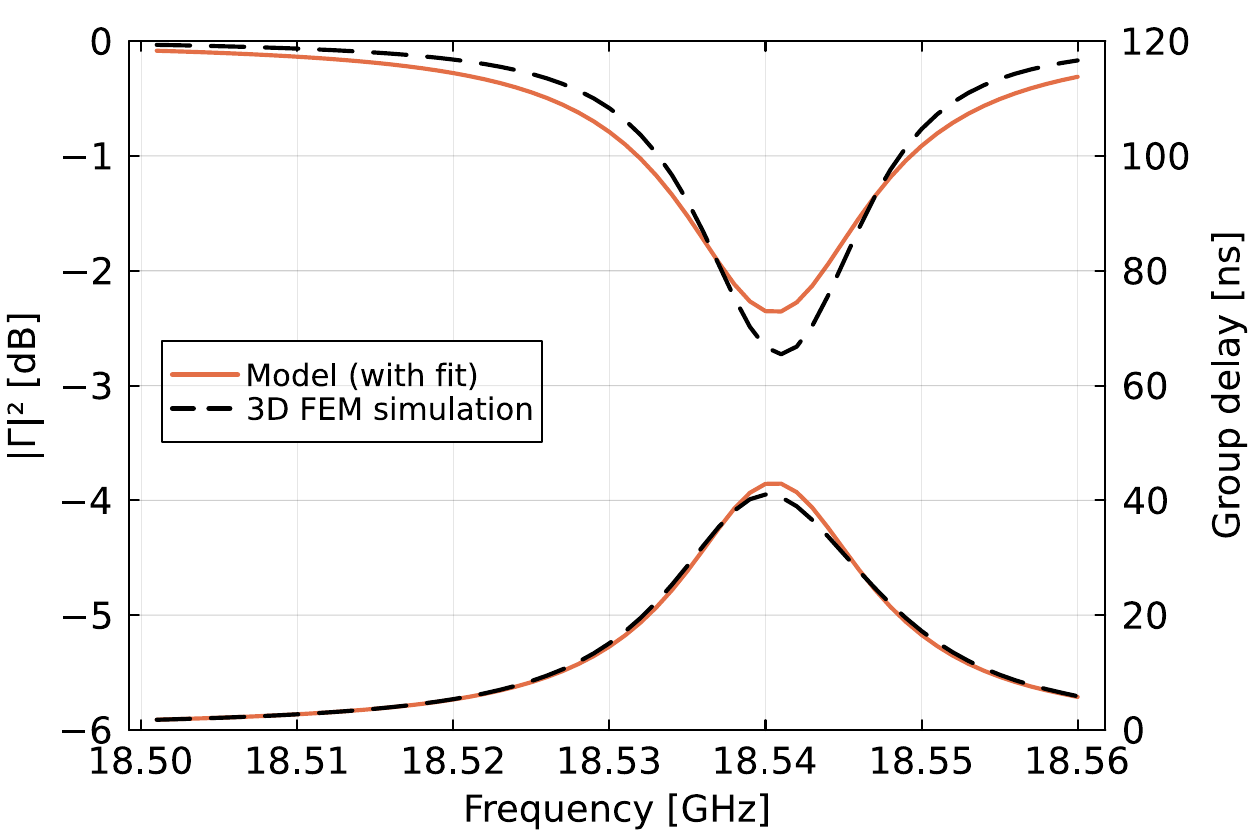}
        \caption{Comparison of reflectivity from 3D FEM simulation and booster model.}
    \end{subfigure}\\
    \vspace*{0.5em}%
    \begin{subfigure}[t]{\textwidth}
        \centering\hspace*{-1.3cm}\includegraphics[width=0.74\textwidth]{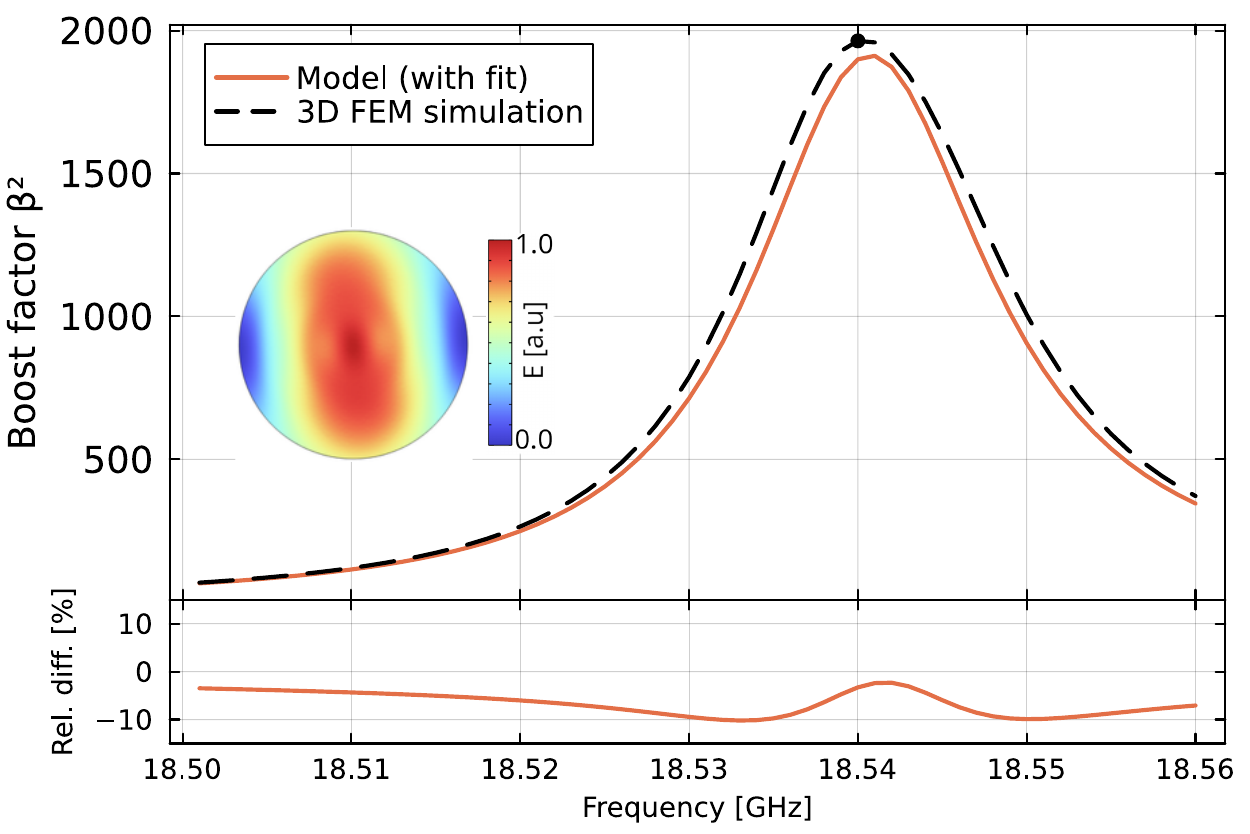}
        \caption{Comparison of boost factor from 3D FEM simulation and booster model.}
    \end{subfigure}
    \caption{Comparison between reflectivity and boost factor obtained from the booster model (orange solid) and the 3D FEM simulation (black dashed) of a three-disk booster with a non-planar disk ($\sim\SI{50}{\micro\m}$ min-max) in front of the mirror.
    The spectra of reflected power $|\Gamma|^2$ and group delay $\tau_\text{gd}$ are shown in (a), whereas the boost factor spectra $\beta^2$ together with their relative difference are shown in (b).
    The simulated transverse electric field distribution at the TE$_{11}$ port is shown as an inset of (b) for the frequency of maximum boost factor.}
    \label{fig:S11_vs_boost_disk}
\end{figure}

Deviations of disk planarity (non-uniformity in the thickness across the surface of the disk) also modify the field distribution in a manner not captured by the booster model.
Due to the localized resonance, we again focus on the region between mirror and closest disk.
The measured planarity of that disk~\cite{MADMAX:2024pil} is used to approximate a realistic disk shape within COMSOL with a min-max planarity of $\sim\SI{50}{\micro\m}$.
It is included in the 3D FEM simulation of the ideal booster and boost factor and reflectivity simulations are repeated.
The agreement of the model with the 3D FEM simulation after fitting its parameters is shown in \cref{fig:S11_vs_boost_disk}.
While the model does not fully reproduce the reflected power and group delay at the frequency of the booster mode, the modeled boost factor curve still closely follows the simulation and their maxima agree within $\SI{5}{\percent}$.
This also highlights the robustness of the booster model: although it does not reproduce the reflectivity exactly, it still manages to determine the boost factor within an acceptable uncertainty.
Interestingly, the boost factor is underestimated in this case, showing that 3D effects can also lead to an increase in sensitivity.

In all cases, the largest deviation from the nominal value is seen in the effective dielectric loss parameter $\tan\delta$, which increased by about one order of magnitude.
It is responsible for reproducing the decrease in boost factor amplitude.
The shift in frequency is reproduced mainly by a deviation of $nd_\text{disk}$ of the order of \SI{10}{\percent} and $d_3$ of the order of \SI{1}{\percent}.
There is no qualitative difference between the effect of the mirror tilt and the effect of a deformed disk in the resulting effective parameters.

This analysis demonstrates that the booster model can account for the effect of small imperfections and 3D effects through the adjustment of effective disk positions, thicknesses, permittivities and losses.
For the studied booster, the minimum value of $|\Gamma|^2$ at resonance is a reliable indicator for the deviation of the setup from the ideal case.
Resonances with the minimum of $|\Gamma|^2$ above \SI{-5}{\decibel} are well reproduced by the model.
This behavior is expected to be largely independent of the booster’s transverse size, given a well-separated booster mode, which will be discussed below.
The difference between simulated and modeled boost factor is by far dominated by a change in amplitude, visible by deviations of boost factor maximum of up to \SI{10}{\percent}, as opposed to frequency.
The study therefore confirms the uncertainty on $|\eta_A|^2$ of \SI{12}{\percent} as estimated originally from electric field measurements~\cite{ary_dos_santos_garcia_first_2025}.
		

\subsection{Control of higher-order modes}
\label{sec:higher-order-modes}

\begin{figure}
	\centering
	\includegraphics[width=0.7\linewidth]{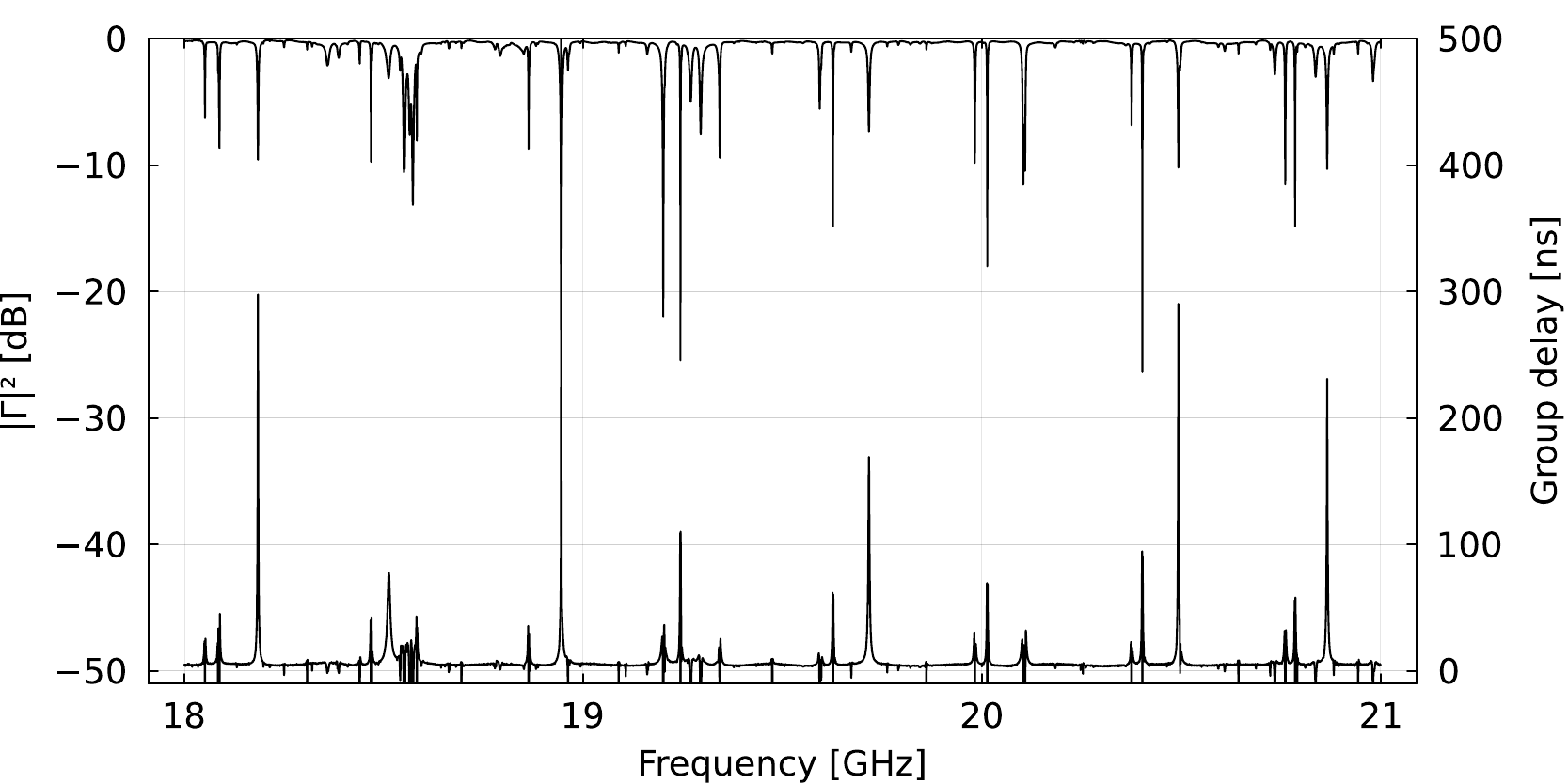}
    \caption{Exemplary reflectivity measurement showing reflected power $|\Gamma|^2$ (top curve) and group delay $\tau_\text{gd}$ (bottom curve) of CB200 over the full range of the taper, demonstrating the highly overmoded behavior of the booster.}
	\label{fig:cb200-refl}
\end{figure}

\begin{figure}
		\centering
		\includegraphics[width=1\textwidth]{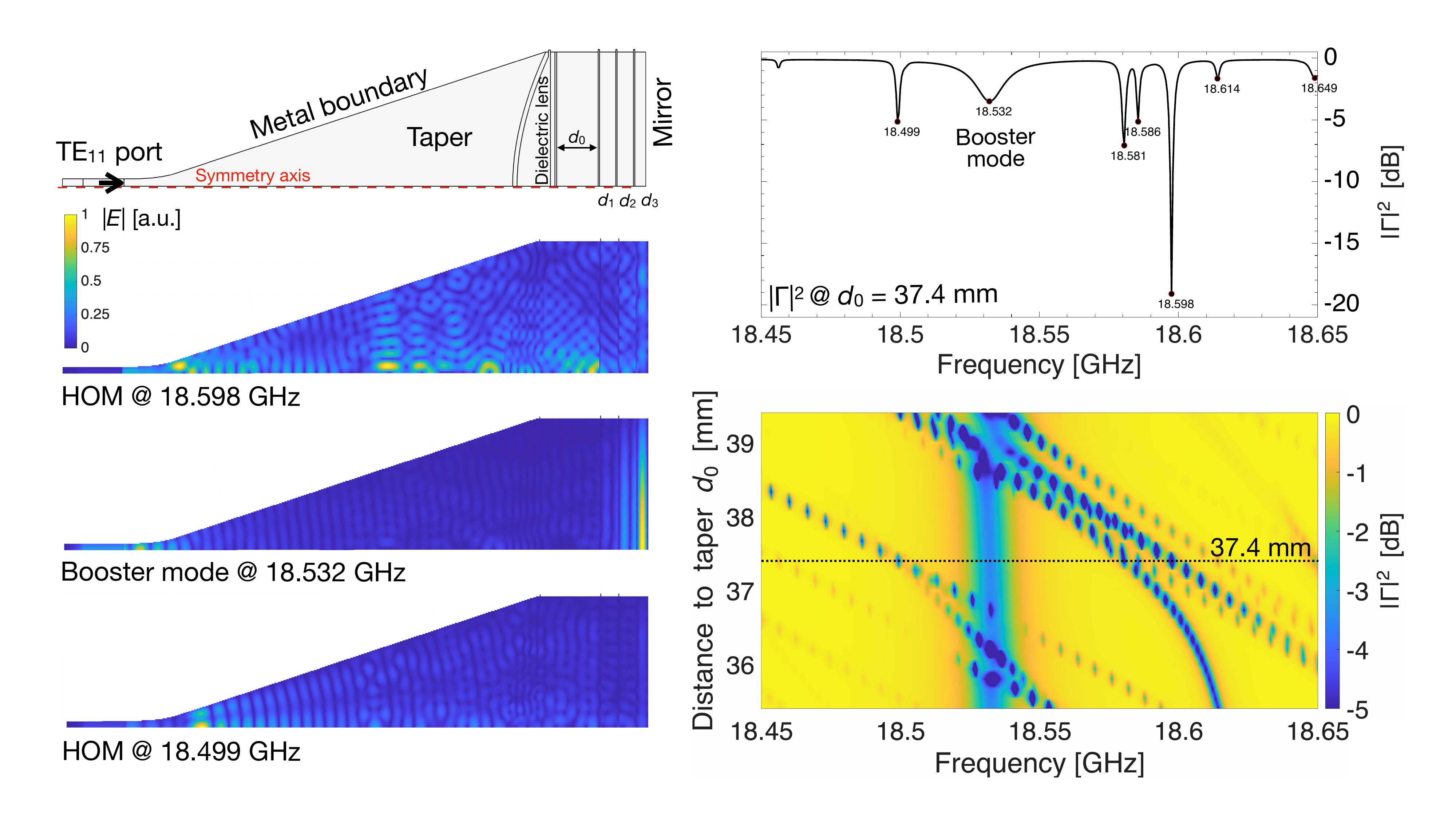}
		\caption{Axis-symmetric simulation of CB200 including the taper obtained for the booster configuration of run~1.1. (Top-left) Illustration of the geometry and boundary conditions of the simulated CB200 setup. The taper input allows only for TE$_{11}$ propagation.
        (Bottom-left) Normalized E-field magnitude shown for the booster mode and adjacent HOMs at the indicated frequencies.
		(Top-right)  Simulated reflectivity spectrum for $d_0 = \SI{37.4}{\mm}$, highlighting the booster mode and nearby HOMs.
        (Bottom-right) $|\Gamma|^2$ as obtained by a continuous sweep of the distance to taper $d_0$. }
	 	\label{fig:modemap}
\end{figure}

The booster gains sensitivity with increasing radius.
A key challenge is that this also allows propagation of an increased number of HOMs.
These can be excited by imperfections and taper-induced mixing, then reflect within the volume and produce many additional resonance lines. An example of this overmoded behavior is illustrated in \cref{fig:cb200-refl}.
Simulating this behavior accurately requires the use of the exact taper dimensions within the full-wave simulation.
Performing a full 3D simulation becomes unfeasible at those scales and a 2D axis-symmetric simulation, as depicted in the top left of \cref{fig:modemap}, is performed instead.

To justify the single-mode approximation of the booster model, the booster mode needs to be separated from HOMs in frequency.
To illustrate our strategy, we include the results of a realistic simulation of the taper connected to the booster in \cref{fig:modemap}.
As can be seen from the included field maps (bottom left), the HOMs excited by the taper are mostly standing waves localized around the taper.
Their frequencies are sensitive to the distance of the first disk to the taper aperture $d_0$.
In contrast, the electric field of the booster mode is mostly confined to the region containing the disks, determined by the disk spacings $d_1, \ d_2, \ d_3$, and remains largely unaffected by $d_0$.
The modal map at the bottom-right clearly demonstrates this decoupling: the booster mode is constant at  $\sim\SI{18.53}{\GHz}$, whereas the HOMs vary in frequency, depending on the distance of the leftmost disk to the taper aperture $d_0$.
The separation ring between taper and closest disk is therefore carefully chosen for each configuration to maximize the isolation between booster mode and HOMs in frequency.

Our tuning strategy was extensively tested and it is found to be very robust: in the case of CB200, a typical separation of HOMs of \SIrange{50}{100}{\MHz} is achieved across the taper’s full operational range of \SIrange{18}{21}{\GHz}.
This separation justifies the single-mode approximation of the booster model and even allows further fine-tuning of the booster mode by externally offsetting the mirror.
In this way, a scanning of $\sim\SI{50}{\MHz}$ can be achieved without disassembling the booster and exchanging the separation rings.

\subsection{Identification of the booster mode}
\label{sec:beadpull}

Full-wave simulations of the booster are not practical for identifying the booster mode.
As already mentioned, they fail to fully reproduce the measured reflectivity spectrum accurately because small imperfections and geometric details, such as those found in the dielectric lens, are particularity difficult to model. To unambiguously identify the booster mode, we measure the electric field distribution inside the booster volume.
We employ a non-resonant, perturbation-based field mapping technique \cite{Perturbation_theory_1966}.
In this approach, $|\mathbf{E}|$ is locally inferred from changes in the measured reflectivity caused by a small dielectric object (bead) translated across the booster cross-section.
The perturbed response is resolved by a VNA across the frequency range of interest, enabling direct correlation between the resonance lines in the reflectivity spectrum and their corresponding field profiles.

\begin{figure}
    \centering		\includegraphics[width=1\linewidth]{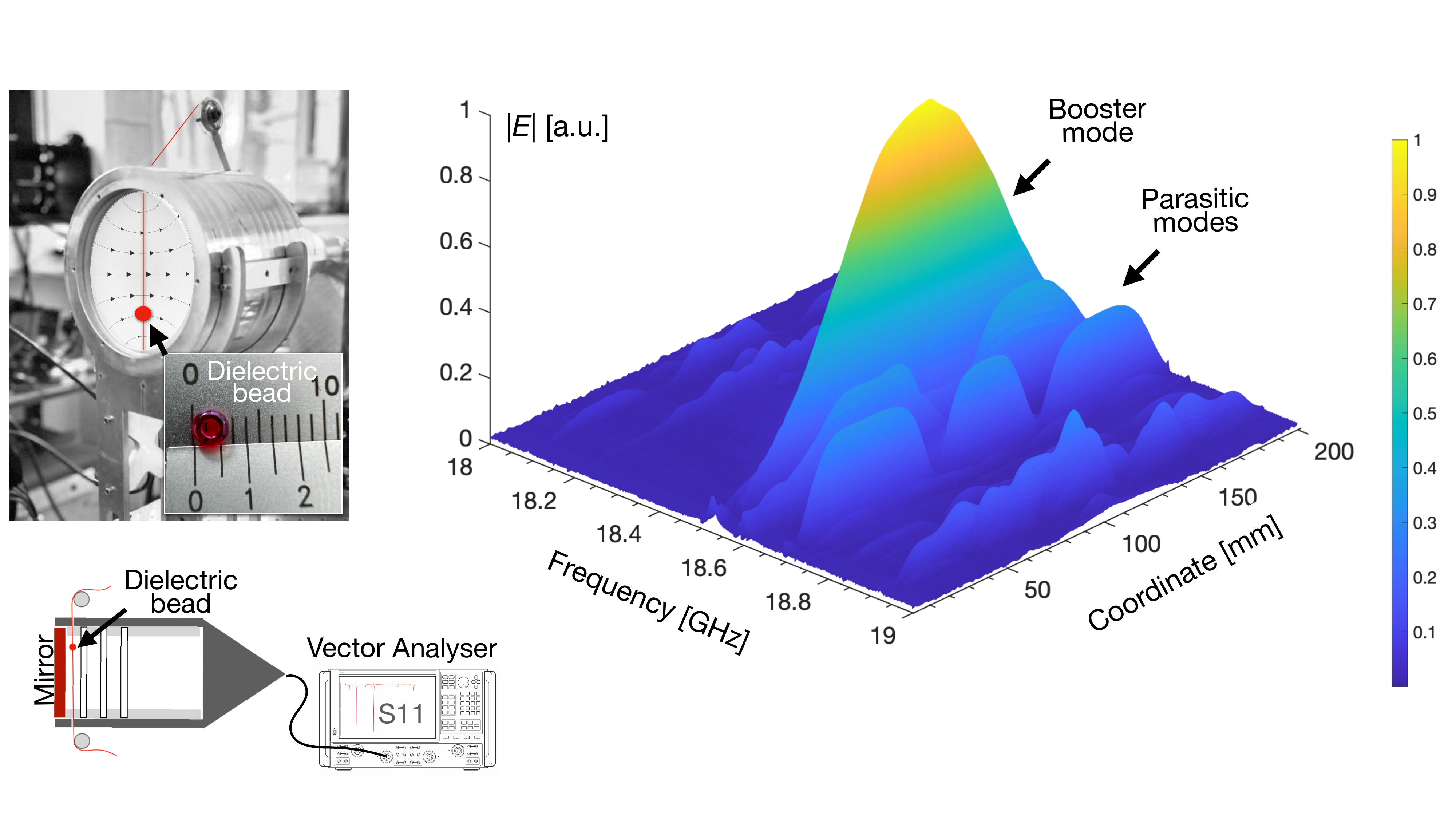}
    \caption{Perturbation-based field measurement implemented for the case of CB200, run 1.1. (Left)  The E-field near the mirror surface is probed with a dielectric bead in the closed booster accessed by two sub-wavelength holes. The field is recorded as a function of the bead positions, indicated by the red line, using a VNA. The contour lines of the desired TE$_{11}$  distribution are overlaid on the photo. (Right) $|\mathbf{E}|$ measured along the line as indicated on the left of CB200 in the \SIrange{18}{19}{\GHz} range. The normalized field clearly reveals the radial structure of the modes. The booster mode appears as a well-defined parabolic shape consistent with the TE$_{11}$ mode. Taken from supplementary material of~\cite{ary_dos_santos_garcia_first_2025}.}
	\label{fig:bpsetup}
\end{figure}

The setup is illustrated in \cref{fig:bpsetup}, along with an exemplary field distribution measured for the CB200 configuration of run 1.1.
The booster mode is distinguished by its lowest-order radial variation, characteristic for the TE$_{11}$ distribution.
In this qualitative approach, the field measurement is sufficient for identifying the booster mode without requiring absolute calibration, unlike~\cite{MADMAX:2024jnp,Egge:2023cos}.

\section{Sensitivity of a closed booster axion search}
\label{sec:data-taking}
The CB200 MADMAX prototype, as shown in \cref{fig:MADMAX}, was used to
take data at CERN within the \SI{1.6}{\tesla} Morpurgo dipole magnet.
Two different disk configurations were used, as described in the upper part of \cref{tab:booster_model}.
The configurations correspond to a maximum sensitivity to axion DM at around \SI{18.55}{\GHz} and
\SI{19.21}{\GHz} respectively.
By fine-tuning the mirror position, the boost factor maximum was shifted by $\mathcal{O}(\SI{10}{\mega\Hz})$, enabling three physics-runs for configuration 1 and two in configuration 2~\cite{ary_dos_santos_garcia_first_2025}.
In this section it is described how the procedure is applied and how uncertainties are determined.

\subsection{Receiver chain characterization}
\label{sec:rec-cal}%
The power calibration of the receiver chain is performed as detailed in
\cref{sec:power-cal} using a calibrated noise source with an excess-noise-ratio (ENR) of $\sim\SI{15}{\decibel}$.
The resulting calibration factors $C$ (see \cref{eq:calfactor}), converting uncalibrated power measurements to system temperature, are included in \cref{fig:cal_factor} in the \cref{sec:power-cal}.
For runs 1.2 and 1.3 the same power calibration was used and therefore an identical $C$ is applied.

The uncertainty of $C$ of around $\SI{10}{\percent}$ is dominated by the uncertainty on the power output of the noise source and the attenuation factor of the attenuator connected between the noise source and the receiver.
Further sources of uncertainty come from the physical temperature and the noise measurements themselves.
The noise source includes a calibration certificate specifying frequency dependent ENR and associated uncertainty.
The values are interpolated linearly between the given frequency points.
For the attenuator, an attenuation of $A = \SI{20 \pm 0.3}{\decibel}$ was approximated by the specified min-max difference of $\sim\SI{1}{\decibel}$.
The uncertainty on the physical temperature is conservatively assumed to be \SI{5}{\kelvin}.
The noise measurements are filtered by a third-order Savitzky-Golay filter~\cite{savitzky_smoothing_1964} with a
$\sim\SI{7}{\mega\Hz}$ window size.
The uncertainty $\sigma_N$ on the noise measurements is estimated by the mean squared error of the $n$ bins of
measurement $N$ and filtered measurement $N_\text{sg}$:
\begin{equation}
    \sigma_N = \sqrt{\frac{1}{n-1}\sum^n_i |N_{i,\text{sg}} - N_i|^2}.
\end{equation}
All uncertainties are treated as uncorrelated and propagated into $C$.

The receiver chain noise is then modeled as described in \cref{sec:rec-noise}.
The main source of parameter uncertainty stems from the power calibration.
Since its uncertainty is almost fully correlated over frequency, a fit to the mean and $\pm 1\sigma$ band of the system temperature spectra is performed.
This results in three sets of parameters that are then interpreted as mean and $\pm 1\sigma$ of the parameters themselves, defining a normal distribution for each model parameter.

The noise parameters determined by the fit are listed in \cref{tab:noise-model} in the \cref{sec:params}.

\subsection{Boost factor determination}
\label{sec:bf-det}

\begin{figure}
    \centering
    \includegraphics[width=0.8\textwidth]{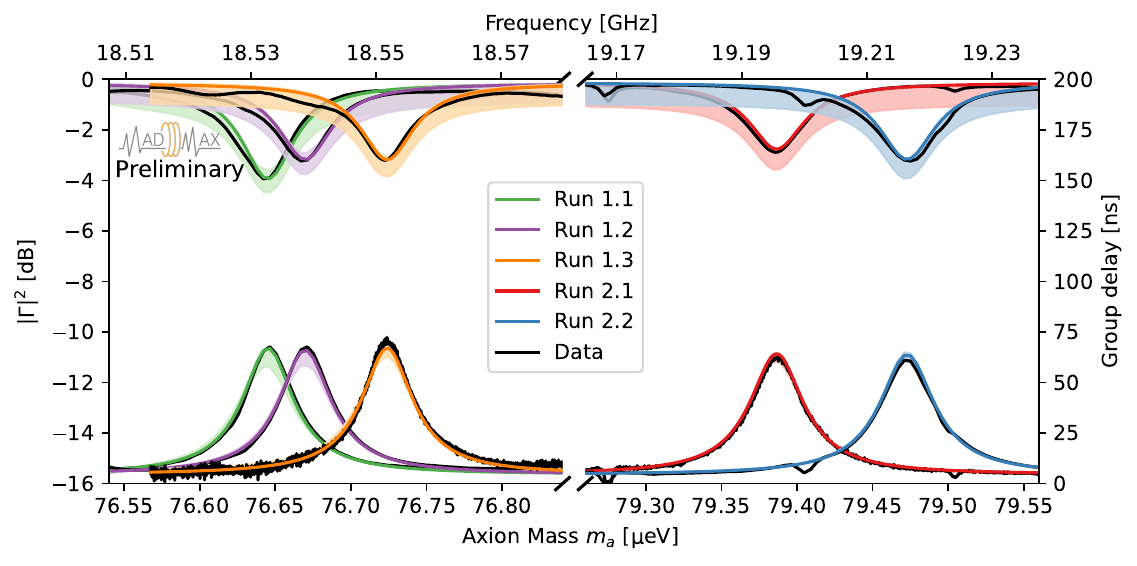}%
    \caption{Reproduction of reflectivity measurements of the five runs of CB200 at the Morpurgo magnet (see \cref{tab:booster_model}).
    The top curves show the reflected power $|\Gamma|^2$, while the bottom curves display the group delay $\tau_\text{gd}$ as a measure of phase.
    The band displays the uncertainty stemming from the fit procedure.
    The feature corresponding to the booster mode is reproduced within uncertainties for all runs.}
    \label{fig:refl_te11}
\end{figure}

After verifying the frequency of the booster mode using the method demonstrated in \cref{sec:beadpull}, the booster model is fit to the measured reflectivities of the different booster setups, as described in \cref{sec:booster}.
\Cref{fig:refl_te11} shows the agreement between modeled and measured reflected power $|\Gamma|^2$ and group delay $\tau_\text{gd}$ within the frequency range of the booster mode.
The model reproduces all measurements within uncertainties, with the reflected power $|\Gamma|^2$ consistently being above $\SI{-5}{\decibel}$.
This confirms the model's assumption of a taper that is strongly coupled to the booster mode and is within the range for which the model was found to correctly predict the boost factor within the assigned uncertainty in \cref{sec:3d-effects}.

The resulting parameters and their fit uncertainties are listed in \cref{tab:booster_model}.
The biggest deviations between fitted and nominal value are seen for the effective dielectric loss parameter $\tan\delta$, which is one order of magnitude higher than the dielectric loss of sapphire, as well as for the disk phase thickness $nd_\text{disk}$, which is about $\SI{25}{\percent}$ higher than the nominal value.
Since these parameters reflect the effective treatment of 3D effects, it is not expected that they match with the nominal values.
For error propagation, a multivariate normal distribution is assumed, which is defined by the covariance matrix estimated from the fit result.
The error band in \cref{fig:refl_te11} reflects the mean and $\pm1\sigma$ of the fit uncertainty.

\begin{table}[t]
    \centering
    \begin{tabular}{l c c c c c}
         & $d_1$ [mm] & $d_2$ [mm] & $d_3$ [mm] & $nd_\text{disk}$ [mm] & $\tan\delta$ \\\toprule
        Config 1 & \num{12.52 \pm 0.01} & \num{12.25 \pm 0.01} & \num{8.38 \pm 0.01} &
        \num{3.06 \pm 0.04} & \num{1e-5}\\
        Config 2 & \num{11.89 \pm 0.01} & \num{12.25 \pm 0.01} & \num{8.02 \pm 0.01} &
        \num{3.06 \pm 0.04} & \num{1e-5}\\
        \midrule
        Run 1.1 & \num{12.2 \pm 0.6} & \num{12.14 \pm 0.08} & \num{8.0979 \pm 0.0002} &
        \num{4.045 \pm 0.003} & \num{4.8 \pm
        0.2e-4}\\
        Run 1.2 & \num{12.2 \pm 0.6} & \num{12.14 \pm 0.07} & \num{8.0953 \pm
        0.0002} & \num{4.043 \pm 0.003} &
        \num{3 \pm 0.1e-4}\\
        Run 1.3 & \num{12.1 \pm 0.1} & \num{12.13 \pm 0.02} & \num{8.08954 \pm
        0.00004} & \num{4.0402 \pm 0.0007} &
        \num{3 \pm 0.03e-4}\\
        Run 2.1 & \num{11.72 \pm 0.05} & \num{11.753 \pm 0.007} & \num{7.80833 \pm
        0.00002} & \num{3.8907 \pm 0.0003} &
        \num{3 \pm 0.01e-4}\\
        Run 2.2 & \num{11.73 \pm 0.07} & \num{11.752 \pm 0.009} & \num{7.81706 \pm
        0.00002} & \num{3.8981 \pm 0.0004} &
        \num{2 \pm 0.01e-4}\\
\end{tabular}
    \caption{Measured (top) and effective model (bottom) parameters. The effective parameters are extracted from a fit to the measured booster reflectivity. $d_1$ to $d_3$ describe the spacings between the disks, $nd_\text{disk}$ is the product of refractive index $n$ and disk thickness $d_\text{disk}$ and $\tan\delta$ the (effective) dielectric loss of the sapphire disks.}
    \label{tab:booster_model}
\end{table}%

\begin{figure}
    \centering
    \begin{subfigure}[b]{0.5\textwidth}
        \centering
        \includegraphics[width=0.95\textwidth]{assets/2024-02-21_overnight_noisefit_paper_cal.pdf}
        \caption{Noisefit of run 1.1.}
        \label{fig:fig-a}
    \end{subfigure}%
    \begin{subfigure}[b]{0.5\textwidth}
        \centering
        \includegraphics[width=0.95\textwidth]{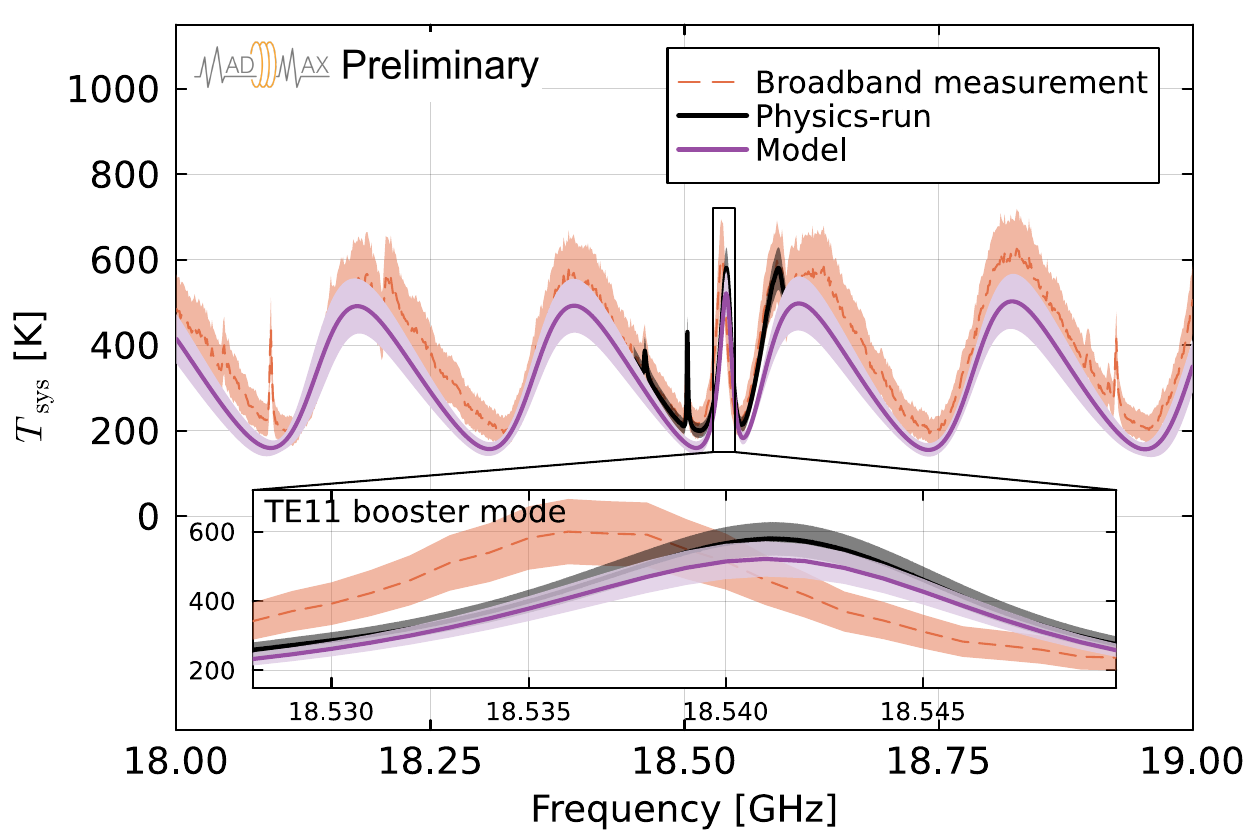}
        \caption{Noisefit of run 1.2.}
    \end{subfigure}\\
    \begin{subfigure}[b]{0.5\textwidth}
        \centering
        \includegraphics[width=0.95\textwidth]{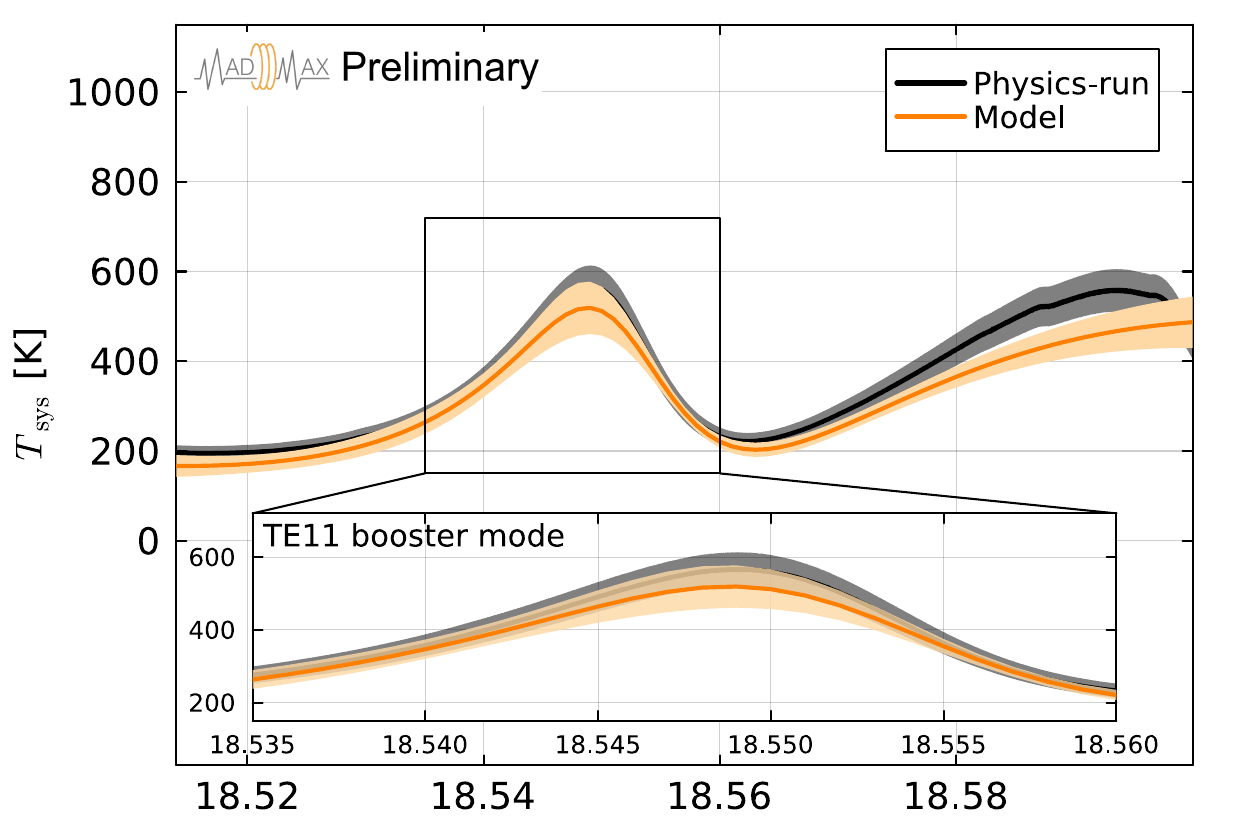}
        \caption{Noisefit of run 1.3.}
    \end{subfigure}%
    \begin{subfigure}[b]{0.5\textwidth}
        \centering
        \includegraphics[width=0.95\textwidth]{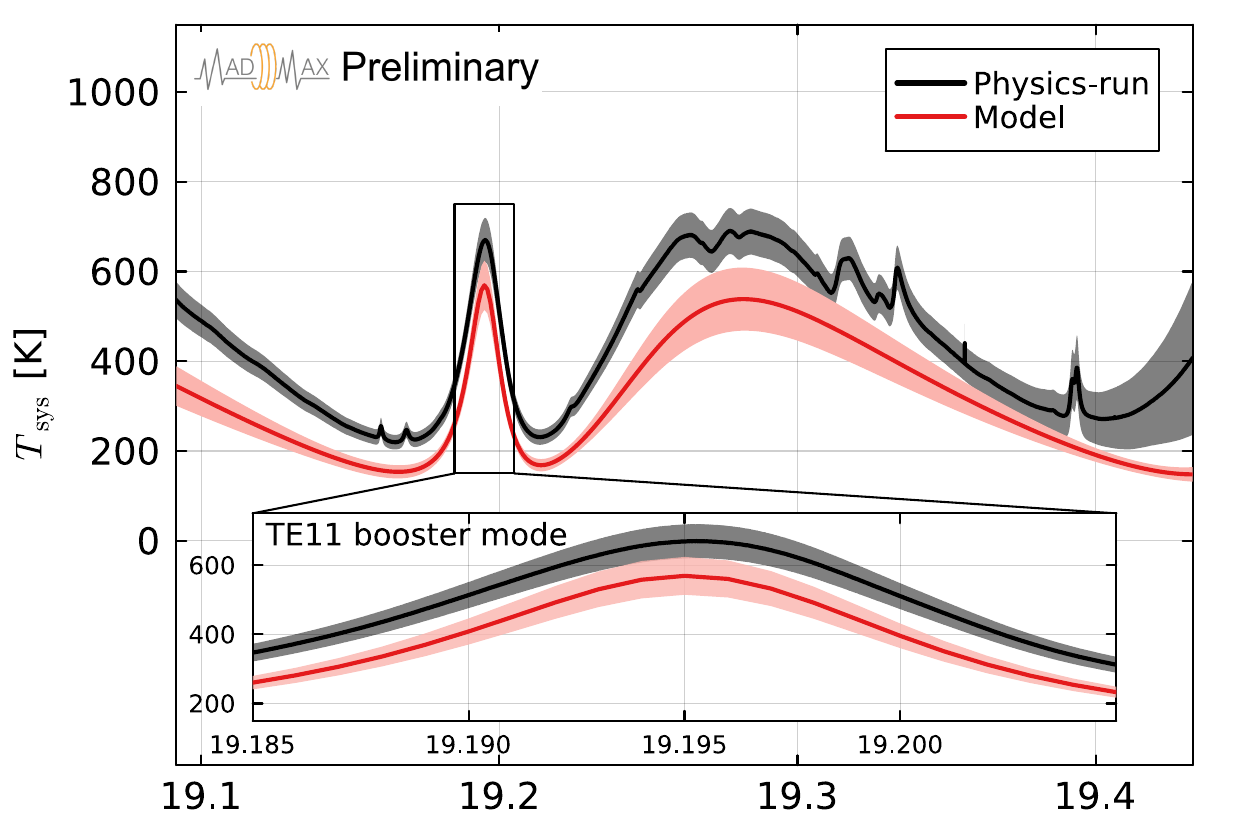}
        \caption{Noisefit of run 2.1.}
    \end{subfigure}\\
    \begin{subfigure}[b]{0.5\textwidth}
        \centering
        \includegraphics[width=0.95\textwidth]{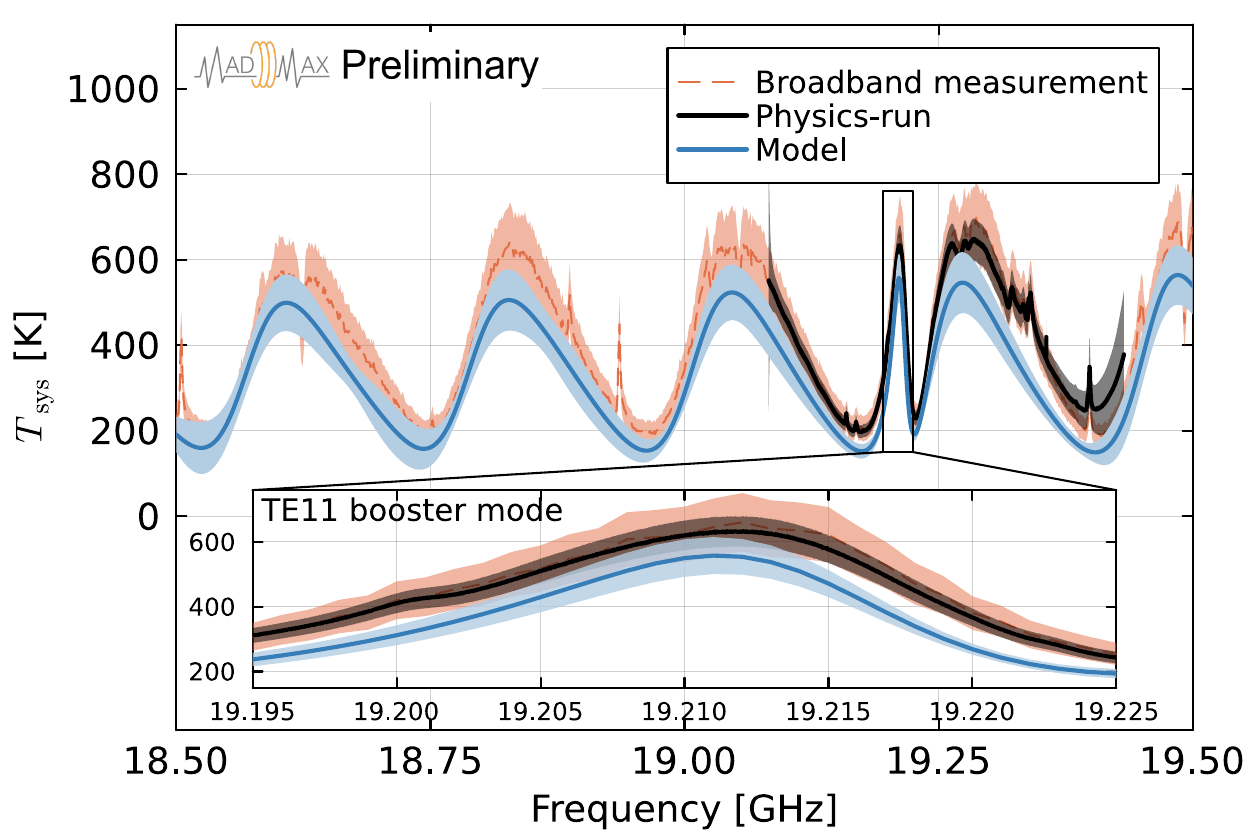}
        \caption{Noisefit of run 2.2.}
    \end{subfigure}%
    \caption{Reproduction of system temperature spectra of the five runs of CB200 at the Morpurgo magnet (see \cref{tab:booster_model}).
        The noise model (different colors) matches the oscillation of the broadband measurement (orange) and the peak position of the booster mode of the physics-run (black), magnified in the inset.
        The bands signify the corresponding uncertainties.
        For runs 1.3 and 2.1, no broadband measurements are available. \Cref{fig:fig-a} has already been shown as \cref{fig:noisefit-example}.}
    \label{fig:noise_fits}
\end{figure}

The system temperature spectra obtained with the combined noise model, using the booster parameters from \cref{tab:booster_model} and receiver noise parameters, as shown in \cref{tab:noise-model} in the \cref{sec:params}, are then fit to the measured system temperature spectra of the different booster setups.
The results are shown in \cref{fig:noise_fits}.
Only the distance between LNA and booster $L_\text{con}$ is left as a free fit parameter.
The uncertainty band stems from propagating the uncertainties of the LNA and booster parameters using a Monte Carlo method, described further in \cref{sec:bf-unc}.
The broad oscillation pattern, stemming from a standing wave between LNA and booster, as well as the resonance peak of the booster mode are reproduced for all setups within uncertainties, demonstrating the capability of the model to simulate the full system.
For two of the runs, no broadband measurement exists.

In both cases, the previous run was performed using the exact same setup with only the mirror position changed using the tuning mechanism.
Therefore, the best fit of the previous run was used as initial value for $L_\text{con}$ in those cases.
The fitted values are listed in \cref{tab:noise-fit} in the \cref{sec:params}.

\begin{figure}[htpb]
    \centering
    \includegraphics[width=0.6\textwidth]{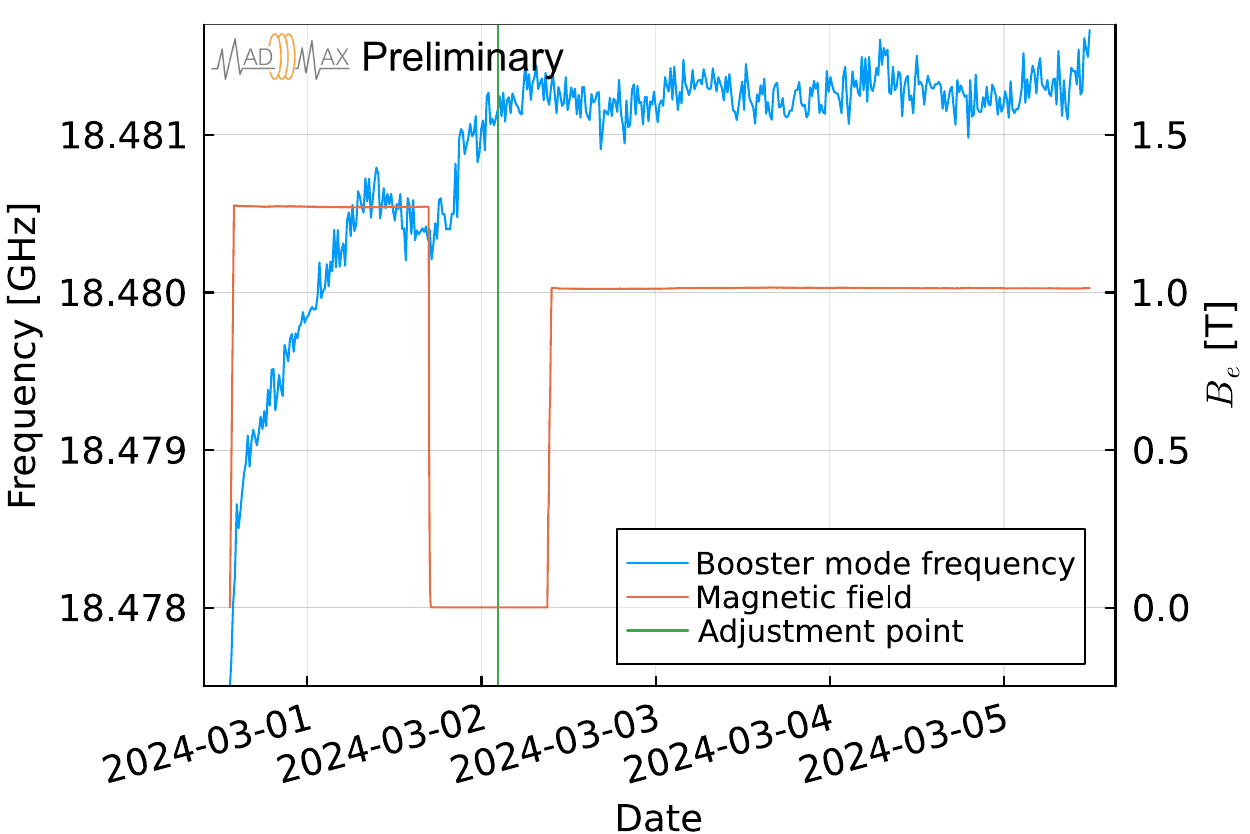}%
    \caption{Booster mode frequency over time of run 1.2 (see \cref{tab:booster_model}) magnet ramp up/down. The blue curve shows the significant increase of the resonance frequency over the first day.
    The orange curve displays the magnetic field $B_e$ and the green bar signifies the time at which the first data point used in the analysis was taken.}
    \label{fig:res-freq-shift}
\end{figure}

In some cases, the booster mode resonance frequency had changed significantly over time, as evident in run 1.2.
\Cref{fig:res-freq-shift} shows its evolution over the period of data taking, with the booster mode frequency stabilizing only after a second ramp up of the magnet had finished.
The main source of mechanical instability within the system is the mirror tuning mechanism, which is likely to take some time to stabilize after the initial setup.
The only change apparent in the system temperature spectrum was a change of the resonance position of the booster mode, which mainly depends on the mirror position.
In contrast, no change of frequency of higher-order-modes, that mainly depend on the distance of the first disk to the taper, is observed.
It is therefore concluded that the change in resonance frequency is caused by a mechanical change of the mirror position.
An additional single parameter fit of the mirror position within the combined booster and receiver model is performed to reproduce the frequency of the booster mode resonance in the system temperature spectrum, resulting in $\mathcal{O}(\si{\micro\m})$ changes from the initially determined parameter.
The modeled spectra shown in \cref{fig:noise_fits} already include this readjustment, which causes them to always match the resonance frequency of the physics-run, not necessarily the broadband calibration measurement.
Data taken before the point of readjustment were discarded in the further analysis.

For the remaining time of the data taking the resonance frequency did not change significantly, with the standard deviation of the peak position being of $\mathcal{O}(\SI{100}{\kHz})$.
The variation is determined for each run $r$ individually and is assigned a normal distribution $\Delta F_r$ given by the mean and standard deviation of the frequency of the booster mode resonance, which is propagated to $\beta^2$ as uncertainty.

\subsection{Uncertainty propagation}
\label{sec:bf-unc}

The total uncertainties on $\beta^2$ are dominated by the uncertainties of the booster model parameters and the field shape uncertainty.
The field shape uncertainty of $|\eta_A|^2 = \num{0.84 \pm 0.1}$ is estimated from electric field measurements and confirmed by the simulations described in \cref{sec:model-verification}.
The uncertainties of the booster model parameters are given by a multivariate normal distribution estimated by the covariance matrix of their fit result.
Further considered are the uncertainties of the receiver noise model parameters, given by individual normal distributions as estimated from the power calibration uncertainty.
With all model parameters being assigned a distribution, a Monte Carlo method is employed to propagate the uncertainties to the final boost factor:
The fits of $L_\text{con}$ and $d_3$ to the measured system temperature are repeated $N=100$ times, each time fixing all other parameters by drawing a random sample from their assigned distributions.
Each resulting boost factor $\beta_\text{1D}^2$ is then further multiplied by a number drawn from the normal distribution $0.84 \pm 0.1$ to take into account the field shape uncertainty and shifted by a frequency drawn from $\Delta F_r$ to consider the observed frequency instability of the booster mode resonance.
The result is considered a sample of the final boost factor distribution $\beta^2(\nu)$.
This method results in $N$ boost factor spectra, from which the mean and standard deviation are determined for each frequency point.
The mean boost factor varied by $<\SI{1}{\percent}$ over the last 25 samples, confirming convergence.

\begin{figure}[htpb]
    \centering
    \includegraphics[width=0.9\textwidth]{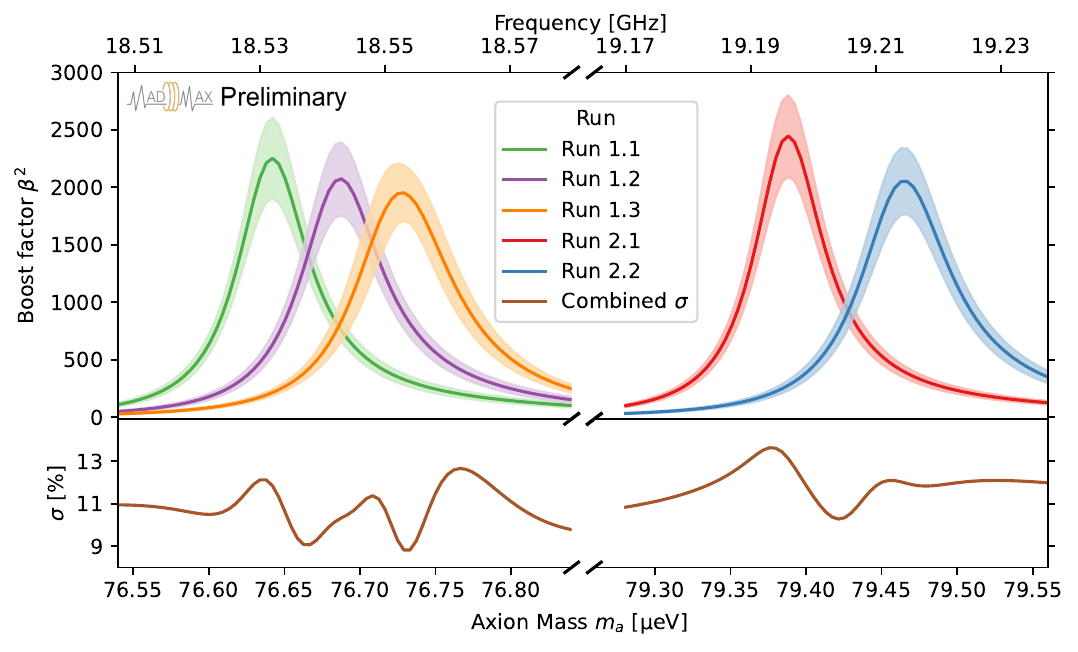}
    \caption{Boost factor distributions for the five runs (see \cref{tab:booster_model}) and their combined relative uncertainty. The error band shows the standard deviation of the individual boost factor samples at each frequency point. The combined uncertainty shown in the bottom panel is the relative uncertainty of the boost factor sum with uncertainties added in quadrature.}
    \label{fig:bf-final}
\end{figure}

For the determination of the sensitivity on $|g_{a\gamma}|$, the boost factors of all datasets are combined, assuming uncorrelated uncertainties.
The combined relative uncertainty is shown in \cref{fig:bf-final}, together with the full boost factor distributions.
The setup achieves boost factors of up to $\sim\num{2500 \pm 300}$ with a combined uncertainty between \SIrange{9}{14}{\percent}, depending on frequency.
With the boost factor distributions in place, a statistical method described in detail in~\cite{diehl_diss}, adapted from~\cite{HAYSTAC:2018rwy}, was used to set limits on the axion parameter space in~\cite{ary_dos_santos_garcia_first_2025}.
\section{Conclusion}
\label{sec:conclusion}

A method to determine the sensitivity of a closed dielectric haloscope to axion dark matter is described in this paper.
A well-known transmission line approach is utilized to model the electromagnetic response of the setup.
By fitting the model parameters, it is able to accurately reproduce the reflectivity of the booster, as measured by a VNA, and the system temperature spectra, as measured by the receiver system.
The model is validated against finite element simulations and is shown to absorb three-dimensional effects such as disk shape deviations and mirror tilt into effective parameters.
It can therefore be used to determine the sensitivity to axion DM of the full setup, consisting of booster and receiver system.
Uncertainties resulting from the power calibration, the fit procedure, the field shape and the time stability of the experiment are considered.
The procedure is applied to data taken with the CB200 prototype at the Morpurgo magnet at CERN~\cite{ary_dos_santos_garcia_first_2025}.

The presented model is able to replace time-consuming and computationally expensive simulations by simple calculations.
The demonstrated ability to reproduce system temperature spectra makes it possible to relate system changes during data taking to specific model parameters.
With this capability, unexpected changes of the setup can be adjusted for without interrupting data taking.
Being largely volume independent, the model allows for further upscaling of the experiment.
This work therefore lays the foundation for future dark matter searches by MADMAX, utilizing more complex booster systems with a higher number of disks and more extensive tuning capabilities.

\acknowledgments

The authors would like to thank the CERN central cryogenic laboratory and the CERN magnet team for the support during all the measurements with the Morpurgo magnet. We thank O.~Reimann, D.~Strom, and A.~Hambarzumjan for significant contributions during the first phase of the MADMAX project. We thank O.~Rossel for providing the drawing of the setup and artistic support. We acknowledge support by the Helmholtz Association (Germany); Deutsche Forschungsgemeinschaft (DFG, German Research Foundation) under Project No. 550641633 and Germany’s Excellence Strategy—EXC 2121 “Quantum Universe”—390833306. This work was also supported by the co-funded project HALOX between Agence national de la Recherche (ANR) and DFG grant (ANR-24-CE92-0049, DFG project No545526387). Computations were performed with computing resources granted by RWTH Aachen University under Project No. 7936. S.~Knirck is supported by Fermi Research Alliance, LLC under Contract No. DEAC02-07CH11359 with the U.S. Department of Energy, Office of Science, Office of High Energy Physics. We acknowledge the support of the MADMAX project by the Max Planck Society.

\appendix
\section{Appendix}

\subsection{Receiver power calibration}
\label{sec:power-cal}
To take measurements of system temperature that are then reproduced by the noise model, the receiver system needs to be calibrated.
The goal of this calibration is to convert power measurements performed with the receiver system to system temperature $T_\text{sys}$ - that is the temperature at which a black body would emit a noise power equivalent to that of the connected device together with the receiver.

The components of the receiver system are shown at the bottom of \cref{fig:MADMAX}.
The booster is connected via a coaxial connection to a chain of amplifiers and filters in front of a spectrum analyzer (SA) to digitize the signal.
The amplification is required, since the expected signal power is below the noise floor of a typical SA.
The bandpass filters are added to avoid saturation of the amplifiers.
A cavity filter at the end of the chain prevents image frequencies, mixer sidebands and aliasing effects from entering the data.

In practice, calibrating the receiver means determining its frequency dependent gain and added noise.
This is done by connecting it to a calibrated noise diode, which is matched to the
\SI{50}{\ohm}  coaxial connection between the receiver and the diode.
Two measurements are performed, $N_\text{on}$ and $N_\text{off}$, one with the noise source
switched on and one with it switched off.
It is assumed that the receiver chain behaves linearly within this range.
The power emitted by the noise source $P_\text{N}$ is given by the Rayleigh-Jeans approximation of Planck's law, that is
\begin{align}
    P_\text{N,off} &= k_B B T_0\label{eq:P_noff}\\
    P_\text{N,on} &= k_B B\;10^{\text{ENR}/10}T_0,\label{eq:P_non1}
\end{align}
with the Boltzmann constant $k_B$, the resolution bandwidth of the receiver $B$, the physical
temperature $T_0$ and the excess noise ratio (ENR) of the noise source.
To not saturate the receiver chain, an attenuator of $A = \SI{20}{\decibel}$ is added between receiver chain and noise source, resulting in a modification to \cref{eq:P_non1}:
\begin{equation}
    \label{eq:P_non}
    P_\text{N,on} = k_B B \left[10^{(\text{ENR}-A)/10}T_0 + (1 - 10^{-A/10})T_0\right] \equiv k_B
    B T_\text{N},
\end{equation} where the term in square brackets is the sum of the attenuated noise source's
power and the thermal noise emitted by the attenuator itself.\footnote{From this, \cref{eq:P_noff} can be recovered by setting $\text{ENR}=0$ and \cref{eq:P_non1} by setting $A=0$}
The receiver chain also adds noise to the measurements, quantified by its equivalent
noise temperature $T_\text{e}$.
It is assumed to be fully uncorrelated to the noise emitted by the noise source, which would result in a simple sum for the total power $P_\text{on/off} = k_B B T_e + P_\text{N,on/off}$.
In the present case, however, the receiver chain is not perfectly matched to the \SI{50}{\ohm}
coaxial connection, resulting in some of the noise source's power being reflected.
Therefore, the actual total power delivered to the receiver chain is~\cite{pozar2011microwave}
\begin{equation}
    \label{eq:P_on/off}
    P_\text{on/off} = k_B B T_\text{e} + (1 - |\Gamma_\text{RC}|^2)k_B B T_{\text{N}/0},
\end{equation} with the receiver's input port reflection given by $\Gamma_\text{RC}$.
The factor can be understood intuitively by thinking about the reflected power corresponding to
$|\Gamma_\text{RC}|^2$ being subtracted from the total power.

Note that this simple expression is only possible due to the fact that the noise diode itself is matched to the coaxial connection, which is not true for the booster.

For the matched noise diode, the receiver chain gain $G$ is given by the measured noise
\begin{equation}
    N_\text{on/off} = G\ P_\text{on/off},
\end{equation}
which, by inserting \cref{eq:P_on/off}, can be written as
\begin{align}
    \label{eq:N_on}
    N_\text{on} &= Gk_BB \left[T_\text{e} + (1 - |\Gamma_\text{RC}|^2) T_\text{N}\right]\\
    N_\text{off} &= Gk_BB \left[T_\text{e} + (1 - |\Gamma_\text{RC}|^2) T_{0}\right].\label{eq:N_off}
\end{align}
The term in brackets defines the system temperature $T_\text{sys}$ for a given device-under-test (DUT) with equivalent temperature $T_\text{DUT}$:
\begin{equation}
    \label{eq:Tsys}
    T_\text{sys} \equiv T_\text{e} + (1 - |\Gamma_\text{RC}|^2) T_\text{DUT}.
\end{equation}
Solving \cref{eq:N_on} and \cref{eq:N_off} for $G$ and $T_\text{e}$ results in
\begin{align}
    G &= \frac{N_\text{on} - N_\text{off}}{P_\text{on} - P_\text{off}}\\
    T_\text{e} &= \frac{N_\text{on/off}}{G k_B B} - (1 - |\Gamma_\text{RC}|^2)T_{\text{N}/0}.
\end{align}
The resulting $G$ not only includes the gains and attenuations of the different
filters and amplifiers of the receiver system but also the insertion loss of the mixer and eventual internal gain from the analog to digital converter used to digitize the signal.
The equivalent noise temperature should roughly match the specified datasheet value of the first stage amplifier, since all other noise sources are suppressed by the first stage amplifier's gain.
To then convert any measured power spectrum $N$ to system temperature $T_\text{sys}$, it only needs to be multiplied by the calibration factor
\begin{equation}
    \label{eq:calfactor}
    C = \frac{1}{Gk_BB} = \frac{P_\text{on} - P_\text{off}}{k_B B(N_\text{on} - N_\text{off})}
    = (1 - |\Gamma_\text{RC}|^2)\frac{T_\text{N} - T_0}{N_\text{on} - N_\text{off}}.
\end{equation}

\begin{figure}
    \centering
    \includegraphics[width=0.8\textwidth]{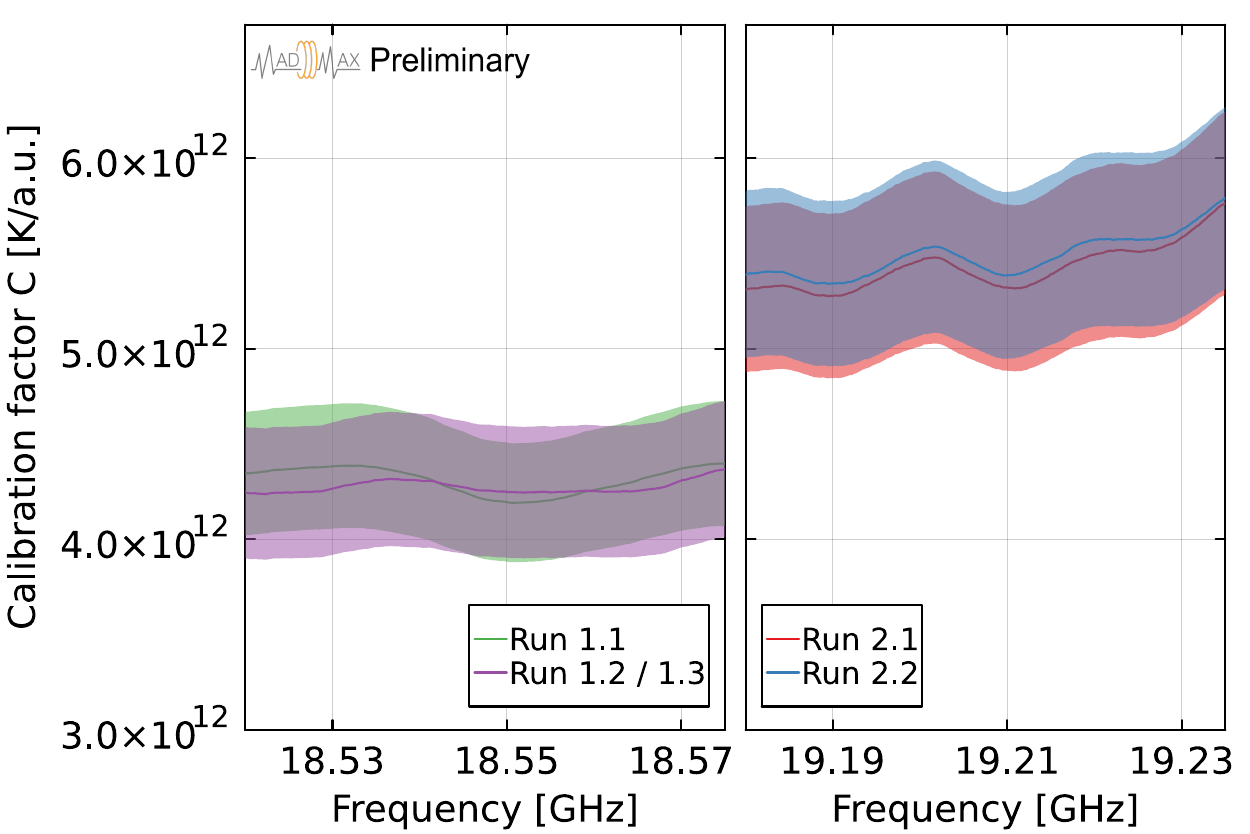}%
    \caption{Calibration factors of the five runs of CB200 at the Morpurgo magnet (see \cref{tab:booster_model}) uncertainties. Three runs were performed around \SI{18.55}{\GHz} (left) and two around \SI{19.21}{\GHz} (right). Note that runs 1.2 and 1.3 use the same power calibration.}
    \label{fig:cal_factor}
\end{figure}

The resulting calibration factors $C$ for the five runs (see \cref{tab:booster_model}) together with their uncertainties are shown in \cref{fig:cal_factor}.

\subsection{Model parameters}
\label{sec:params}

Besides the fitted parameters, the booster model requires a set of fixed parameters listed in \cref{tab:fixed-params}.
$\epsilon_d = n^2$ and $d_\text{disk}$ are included in the fit in the form of their product $nd_\text{disk}$.
$d_0$ and $\sigma_\text{Al}$ are kept fixed throughout.
Since $d_0$ does not affect the boost factor calculation, its uncertainty does not need to be considered further.
While no clear uncertainty on $\sigma_\text{Al}$ was found in the literature, it was checked that a \SI{20}{\percent} variation only changes the boost factor by $<\SI{2}{\percent}$, therefore its uncertainty is neglected.

\begin{table}[b]
    \centering
    \begin{tabular}{c c c c c}
         $L + d_0$ [mm] (C1) & $L + d_0$ [mm] (C2) & $d_\text{disk}$ [mm] & $\epsilon_{d}$ & $\sigma_\text{Al}$ [\si{\siemens\per\m}]  \\\toprule
         \num{203 \pm 1} & \num{187 \pm 1} & \num{1.0 \pm 0.01} & \num{9.36 \pm 0.1} & \num{3.77e7}
    \end{tabular}
    \caption{Fixed parameters used in the booster model for configuration 1 (C1) and configuration 2 (C2).}
    \label{tab:fixed-params}
\end{table}

The fit parameters of the receiver noise model and combined model are shown in \cref{tab:noise-model} and \cref{tab:noise-fit} respectively.

\begin{table}
    \centering
    \begin{tabular}{c c c c c c}
         $V_n$ [pV] & $I_n$ [pA] & $|c|$ & $\phi_c$ & $L_\text{open}$ [ps] & $L_\text{short}$ [ps] \\\toprule
         \num{435 \pm 30} & \num{17 \pm 1} & \num{0.6 \pm 0.1} & \num{358.7 \pm 0.9} & \num{45.8 \pm 0.1} & \num{71.54 \pm 0.08}\\ 
    \end{tabular}
    \caption{Fitted receiver noise model parameters, consisting of voltage and current noise amplitude $V_n$ and $I_n$, their correlation magnitude and angle $c$ and $\phi_c$ as well as the electrical lengths $L_\text{open}$ and $L_\text{short}$ between open and short standard and LNA respectively.}
    \label{tab:noise-model}
\end{table}

\begin{table}
    \centering
    \begin{tabular}{l c c}

         & $L_\text{con}$ [ns] & $d_3$ [mm]\\\toprule
         Run 1.1 & \num{1.58 \pm 0.07} & \num{8.097 \pm 0.001}\\
         Run 1.2 & \num{1.55 \pm 0.05} & \num{8.0916 \pm 0.0005}\\
         Run 1.3 & \num{1.78 \pm 0.47} & \num{8.0882 \pm 0.0006}\\
         Run 2.1 & \num{1.50 \pm 0.04} & \num{7.80833 \pm 0.00002}\\
         Run 2.2 & \num{1.22 \pm 0.04} & \num{7.8155 \pm 0.0001}
    \end{tabular}
    \caption{Fit values for the distance between LNA and booster in the noise model $L_\text{con}$ and the readjustment of the distance between mirror and closest disk $d_3$.}
    \label{tab:noise-fit}
\end{table}

\bibliography{references}
\bibliographystyle{JHEP}

\end{document}